\pdfoutput=1
\documentclass[fleqn,usenatbib,usedcolumn]{mnras}
\usepackage[british]{babel}             
\usepackage{txfonts}                    
\let\la=\lesssim 
\usepackage[T1]{fontenc}                
\usepackage{graphicx}                   

\hypersetup{pdfauthor={A. A. Orsi},
               pdftitle={Clustering of galaxies},
               pdfkeywords={galaxies:high-redshift -- galaxies:evolution -- methods:numerical},
               bookmarksnumbered=true}
\volume{{\rm submitted}}


\newcommand{\mpc}{\mbox{${\rm Mpc}/h$}}
\newcommand{\mhalo}{\mbox{$M_{\odot}/h$}}

\newcommand{\vlos}{v_{\rm los}}

\newcommand{\msg}{LRG}
\newcommand{\sfg}{ELG}

\newcommand{\modc}{Model C}

\newcommand{\twopartdef}[4]
{
	\left\{
		\begin{array}{ll}
			#1 & \mbox{if } #2 \\
			#3 & \mbox{if } #4
		\end{array}
	\right.
}

\title[The impact of galaxy formation on RSDs]
{The impact of galaxy formation on satellite kinematics and redshift-space distortions}
\author [Orsi \& Angulo]
{\'Alvaro A. Orsi\thanks{Email: aaorsi@cefca.es} and Ra\'ul E. Angulo \\
Centro de Estudios de F\'isica del Cosmos de Arag\'on, Plaza de San Juan 1, Teruel, 44001, Spain.\\
}
\begin{document}
\maketitle
\begin{abstract}
Galaxy surveys aim to map the large-scale structure of the Universe and use redshift space distortions to constrain deviations 
from general relativity and probe
the existence of massive neutrinos. However, the amount of information that can be 
extracted is limited by the accuracy of theoretical models used to analyze the 
data. 
Here, by using the {\tt L-Galaxies} semi-analytical model run over the MXXL N-body simulation, 
we assess the impact of galaxy formation on satellite kinematics and the theoretical modelling of redshift-space 
distortions. 
We show that different galaxy selection criteria lead to noticeable differences in the radial distributions and 
velocity structure of satellite galaxies.
Specifically, whereas samples of stellar mass selected galaxies feature satellites that
roughly follow the dark matter, emission line satellite galaxies are located preferentially
in the outskirts of halos and display net infall velocities. We demonstrate that 
capturing these differences is crucial for modelling the multipoles of the 
correlation function in redshift space, even on large scales. In particular, we
show how modelling small scale velocities with a single Gaussian distribution leads to
a poor description of the measure clustering. In contrast, we propose a parametrization 
that is flexible enough to model the satellite kinematics, and that leads to and
accurate description of the correlation function down to sub-Mpc scales. 
We anticipate that our model will be a necessary ingredient in improved theoretical descriptions of 
redshift space distortions, 
which together could result in significantly tighter cosmological 
constraints and a more optimal exploitation of future large datasets.
\end{abstract}


\section{Introduction}
The gravitational law and the amount of matter in the Universe are fundamental
aspects of the cosmological model. A key topic in observational cosmology is to
probe these by measuring the growth of structure encoded in the parameter $f
\equiv {\rm d}\ln D/{\rm d}\ln a$, where $D$ is the linear growth factor at the
scale factor $a = (1+z)^{-1}$. The value of the growth rate, $f$, and its
evolution with redshift has been used to constrain departures from General
Relativity and explore alternative gravity theories \citep{guzzo08}.
Additionally, $f$ can also be employed to constrain the presence of massive
neutrinos, which introduce a scale-dependent growth of structure
\citep{beutler14}. 

The most accurate measurements of cosmic growth rate originate from the
observed anisotropy in the two-dimensional correlation function of galaxies.
Distances to galaxies are inferred from their redshift, a combination of the
Hubble expansion and their peculiar velocity. Thus, galaxy clustering along
the line of sight differs from that on the perpendicular direction, the former
being distorted by the velocity field \citep{jackson72, sargent77}. 
This, in turn, depends on the growth
rate. This anisotropy in the clustering, known as redshift-space distortions
(RSDs), is therefore sensitive to the value of $f$.

Early attempts at interpreting the anisotropic clustering signal have shown the
feasibility of the measurement and the power of RSDs as a cosmological probe
\citep[e.g.][]{peacock01,hawkins03}. To date, the best constraints on $f$ were obtained
by analyses of the Baryon Oscillation Spectroscopic Survey (BOSS). The
combination $f\sigma_8$, with $\sigma_8$ the root-mean-square linear
fluctuation of the mass distribution on scales of $8~\mpc$, was measured with a
precision of $\sim 6\%$ at the redshifts $z = 0.38, 0.51,$ and $0.61$
\citep{alam16}. At higher redshifts, the best constrains were placed by VIPERS
with a $\sim 11-20\%$ precision at $z = 0.86$ \citep{delaTorre2017, mohammad17}. Given the
accuracy of the measurements, the constrains are fully consistent with the
predictions of general relativity and massless neutrinos. 

In the near future, a new generation of multi-object spectroscopic surveys,
such as HETDEX, 4MOST, DESI, PFS and eBOSS
\citep{hill08,dejong12,weinberg13,takada14,delubac17}, and the space missions
Euclid and WFIRST \citep{laurejis10,spergel15}, will aim at significantly
improving current measurements. The goal is to map a larger number of galaxies,
over bigger volumes, and at redshifts above $1$. The larger volume will reduce
the statistical errors, and the extended redshift coverage should help
distinguishing different cosmologies and/or gravity models \citep[see,
e.g.][]{linder16}. For an optimal exploitation of these datasets, upcoming
surveys require an accurate modelling of the density and velocity fields as
traced by galaxies, as well as how these are affected by galaxy formation
physics. 

Over the last couple of years, there has been significant progress on improving the theoretical 
models of RSDs for dark matter and halos \citep{scoccimarro04, tinker07, okumura15,wang16,kopp16,bose17}.
As a result, scales much smaller than those described by linear theory are used in cosmological
analysis. For instance, \citet{sanchez16} claims to obtain cosmological constraints with BOSS data 
on scales above $s \gtrsim 20~\mpc$, consistent with the limit found with dark matter simulations 
in \citet{white15}. In the future, advances in numerical simulations, emulators and perturbation theory will likely
improve the theoretical descriptions of clustering, allowing to reach even smaller scales. 

The next frontier in the modelling of RSDs will be to understand the galaxy scale-dependent
bias and their velocity field, specifically satellite kinematics. Currently, there is widespread use of simplified treatments to account for
small-scale velocities
\citep{bianchi15,uhlemann15,bianchi16,zheng17}. 
However, as we will show in this paper, this is inaccurate and not physically motivated, but
instead shaped by galaxy formation processes. 

The role
of galaxy formation is enhanced by the varied selection criteria of future surveys. 
A common target of future surveys are the so-called emission-line
galaxies (ELGs). ELGs are abundant at high redshifts,  and their redshift can be
precisely measured by identifying narrow strong emission lines in their spectra.
However, ELGs are comparatively much less understood than the Luminous Red
Galaxies (LRGs) targeted by BOSS. These objects are highly biased,
abundant galaxies with a similar stellar mass and an accurate redshift
determination, provided by a $4000$\AA{} break spectral feature typical of old
stellar populations \citep{padmanabhan07}. Instead, ELGs are expected to have
lower bias, sample a wider range of stellar masses, and possibly be more
affected by environmental effects such as ram pressure stripping and by
assembly bias.

In this scenario, the upcoming data is presenting a serious challenge for its
optimal interpretation: to significantly improve the theoretical understanding
of the structure of the velocity fields over the widest possible range of scales, and,
in particular, regarding its connection to galaxy formation physics. This is
the problem we address here. 

In this paper we study how the redshift space clustering and kinematics of
galaxies is shaped by galaxy formation processes. For this, we make use of a
semi-analytical model of galaxy formation embedded in a very large N-body
simulation. Therefore, the relation between the location and velocity of a dark
matter structure and the properties of the galaxy it hosts, is a direct
prediction of an {\it ab-initio} modelling of galaxy formation physics. This
includes tracking the evolution from hot gas in halos to detailed
star-formation histories, feedback processes, mergers and instabilities
triggering startbursts, etc. Rather than attempting to predict the correct
clustering of a particular galaxy population, we aim at exploring how different
galaxy populations deviate from a simple prescription, and identify the
ingredients that can help improving the description of galaxies for probing
cosmology incorporating the small-scale regime.

We illustrate the differences in clustering expected between \sfg s and \msg s.
In particular, we show that \sfg s are hosted by lower mass halos and have
lower satellite fractions. Interestingly, these satellites are preferentially
located in the outermost parts of halos and feature a net infall velocity
component. We propose a physically-motivated parametrization of the intra-halo
velocity field that can capture accurately these aspects and show that this is
sufficient to model the first three non-zero multipoles of the correlation
function down to $\sim 1 \mpc$.

This paper is organized as follows: Section 2 describes the tools and strategy
we use throughout this work. Section 3 discusses the overall properties, host
halo mass distributions, number density profiles, and clustering of galaxy
samples. Section 4 compares different descriptions of intra-halo velocities in
terms of their performance in clustering analysis for galaxy surveys. In
Section 5 we discusses applications and possible extensions of our work.
Finally, we summarize our main findings in the conclusions.

\section{methodology}

In this section we discuss the tools and strategy used in this work to explore
the impact of galaxy formation on redshift-space distortions. We start by
describing how we obtain theoretical predictions for the distribution and
properties of galaxies (\S 2.1), and then describe the selection criteria we
employ to define ELGs and LRGs samples (\S2.2). Finally, we move on to
detailing how we estimate the 2-point correlation function and its covariance
matrix (\S2.3). 

\subsection{The joint numerical modelling of galaxies and their dark matter
halos}

The results presented in this work are based on a galaxy formation model
applied over the merger history of dark matter structures extracted from a very
large $N$-body simulation. Before describing these two ingredients separately,
we emphasize that throughout this paper we do not focus on developing a new
galaxy formation model to improve the agreement with observational data. Such
exercise is beyond the scope of this paper. Instead, we make use of a
state-of-the-art galaxy formation model to develop a model description that
incorporates the impact of galaxy formation processes in the clustering of
galaxies. 

The simulation used in this work is the {\it Millennium-XXL} (MXXL) run,
described in full detail in \citet{angulo12}. This is a dark matter only
simulation of $6720^3$ particles over a cubic volume of $27 ({\rm Gpc/}h)^3$.
The particle mass resolution is $m_p = 6.1 \times 10^9 M_{\odot}/h$. The
cosmological parameters are identical to those of earlier Millennium runs, i.e.
a flat cosmology with $h = 0.73, \Omega_m = 0.25, \Omega_{\Lambda} = 0.75, n =
1$ and $\sigma_8= 0.9$.

The galaxy formation used is a variant of the {\tt L-Galaxies} semi-analytical
model presented in \citet{guo11}.  Briefly, the semi-analytical technique
models the growth and evolution of galaxies in a hierarchical universe by
following the evolution of gas, stars and metals throughout the merger
histories of dark matter halos. The main physical processes included are: the
shock-heating of gas in dark matter halos followed by radiative cooling, which forms a
cold gas component that settles into a rotating disk at the bottom of the
gravitational potential well; the subsequent formation of stars, metals and the
chemical enrichment of the gas, all modulated by feedback mechanisms such as
those caused by supernovae and by the energy released by an active galactic
nucleus (AGN); spheroid formation as a result of galaxy merger and disk
instability episodes; the growth of a supermassive black hole in the centre of
each galaxy; and the computation of observed properties by combining the
star-formation histories of individual galaxies with a stellar population
synthesis code. The free parameters of the model are set by matching a 
diverse set of observations that \textit{does not} include RSDs. It is thus
worth emphasizing that our predictions for the velocity field of galaxies
are a result of physical assumptions.

The mass resolution of the MXXL is a factor of $\sim8$ worse than that of the
Millennium simulation \citep{springel05}. Hence, some modifications were needed
to improve numerical convergence with those obtained with the
same variant of the code over the Millennium simulation. Briefly, these
corrections are: i) A galaxy drawn from the Millennium simulation, hosted by a
halo at the same redshift and mass, is placed in newly formed halos (i.e.
those that up-cross the detection threshold for the first time); ii) merger trees
are extended to lower halo masses to account for the effect of minor mergers
that are unresolved in the MXXL. The details, and convergence tests of these
two corrections are discussed in length in \citet{angulo14}. Here we simply
recall that the real-space correlation function of stellar-mass selected
galaxies shows good agreement with that computed from the Millennium
simulation. 

We apply further modifications to improve convergence of the satellite galaxy
population in redshift space. {\tt L-Galaxies} identifies galaxies as type 0
(i.e. central galaxies), type 1 (satellites with a sub-halo), and type 2
(orphans, i.e.  satellites without a sub-halo). The latter type occurs when a
galaxy's sub-halo has vanished at a particular snapshot of the simulation, due
to mass disruption events or resolution effects. In the variant of the model
run over the Millennium simulation, the position and velocity of type 2
galaxies are assigned by tracking the most-bound particle of the last resolved
sub-halo structure.  Unfortunately, due to huge data storage requirements,
there is no particle information stored for the MXXL simulation. The type 2
positions are thus calculated from the last time their sub-halo was identified,
after which point it is assumed that their radial distance to the host halo
centre shrinks as the square root of the elapsed dynamical friction timescale.
Additionally, the velocity of type 2 galaxies are frozen to the last recorded
value for the sub-halo and then its direction randomized relative to the centre
of the main host halo (assuming that these objects lie within the virial radius
$R_{\rm vir}$, see section \ref{sec.gsample}). 

We have checked that these modifications result in better agreements with the
statistics measured from the Millennium simulation. Specifically, in the Appendix
we show that the detailed spatial and velocity distribution of our galaxy
samples converge for scales above $r\approx 100 {\rm kpc/}h$.

\subsection{Construction of the galaxy samples} \label{sec.gsample}

Employing the theoretical galaxy catalogues described in the previous section,
we can build samples that mimic the selection criteria for current and
future surveys. 

The previous generation of cosmological galaxy surveys targeted mostly red and
massive galaxies. To mimic such samples, we rank order our galaxies according
to the predicted stellar mass and select objects with a global number density
$n = 10^{-3}\,h^3{\rm Mpc}^{-3}$.  Future surveys will, however, also target
galaxies with strong emission lines originated by ionizing radiation 
from young massive stars. These stars
are short-lived, thus these samples can be regarded as selected by star
formation rate. Other astrophysical properties, such as the gas-phase
metallicity and ionization parameter of H{\sc ii} regions also determine the
strength of a nebular emission line \citep{orsi14}. However, they
introduce relatively minor dependencies and are neglected here. Therefore, to
mimic samples of \sfg s we simply select galaxies in a ranked ordered list
according to the predicted star formation rate.

We note that by selecting galaxies according to a given abundance, we can 
compare both galaxy samples more directly and also reduce the sensitivity
of our results to possible mismatches between the predicted and observed
stellar mass functions or star formation rates. Additionally, we note that
we consider only the $z=1$ snapshot, motivated by the target redshift of 
upcoming surveys, however, our results can be qualitatively extended to 
any redshift range.



Finally, additional observational constraints, such as colour selections,
contamination fractions and other instrumental features would be necessary to
fully mimic a survey selection. However, we neglect these effects here as our
goal is to illustrate how the main physical features (e.g. stellar-mass vs.
star-formation rate) translate into different clustering properties. The
construction of detailed mock catalogues representing large cosmological
surveys is out of the scope of this paper, but such calculation can be found
elsewhere \citep[see, e.g.][]{orsi10,merson13}.


\subsection{Clustering measurements} \label{sec.clustering}

We characterize the clustering of the galaxy samples using the two-point
correlation function. Since the simulation box is periodic, we can compute the
2-point auto-correlation function $\xi(r)$ using the direct estimator: 

\begin{equation}
 \xi(r) = \frac{DD(r)}{\bar{n}\Delta V(r) } -1,
 \label{eq.xidirect}
\end{equation}

\noindent where $DD(r)$ stands for the number of pairs of objects within a
radial bin centered at $r$ normalized by the total number of pairs, $\bar{n}$
is the number density of objects in the simulation box, and $\Delta V(r)$ is
the volume of a spherical shell centered at $r$.

To estimate errors in the measurements, we construct a covariance matrix using
the sub-sample method. This consists in sub-dividing the full simulation box
into smaller boxes and computing the clustering in each of these. We divide the
simulation into $N_s = 216$ sub-boxes of length 500 $\mpc$.  The covariance
matrix elements are thus estimated as: 

\begin{equation} C_{ij} = \frac{1}{N_s}
\left\{\frac{1}{N_s-1}\sum_k [w_i^k - \langle w_i \rangle][w_j^k - \langle w_j
\rangle]\right\}, 
\label{eq.cov} 
\end{equation} 

\noindent where each $w$ corresponds to a
vector that contains a series of clustering measurements concatenated over a
given scale range. Since now each sub-box is not periodic, we compute the
auto-correlation function using the Landy \& Szalay estimator \citep{landy93}:

\begin{equation}
\xi(r) = \frac{DD(r) - 2DR(r) + RR(r)}{RR(r)},
\end{equation} 

\noindent where $RR(r)$ is the number of pairs of objects from a
random catalogue at distance $r$, and $DR(r)$ is the number of random objects
around data objects. Finally, we scale the resulting covariance matrix with the 
inverse of the volume of interest \citep{klypin17}.

Throughout this work, we focus on the anisotropic clustering signal that
results from the line-of-sight velocities of galaxies distorting their apparent
redshift. We assume the distant observer approximation, and define the
line-of-sight redshift-space coordinate $s$ as: 

\begin{equation}
 s = r_z + \frac{v_z}{aH(z)},
\end{equation}

\noindent where $r_z$ is an arbitrary cartesian coordinate representing the
line-of-sight direction, $v_z$ is the peculiar velocity component along this
direction, $a = (1+z)^{-1}$ and $H(z)$ is the Hubble parameter at redshift $z$.

The anisotropic redshift-space correlation function is obtained by computing the
two-point correlation function as a function of perpendicular $r_{\perp}$ and
parallel $r_{\parallel}$ positions, $\xi(r_{\perp},r_{\parallel})$. This can be
expressed as a multipole expansion in Legendre polynomials $L_{\ell}(\mu)$.
Each multipole term of order $\ell$ is computed as:

\begin{equation}
\xi_{\ell}(s) = -\frac{2\ell + 1}{2} \int_0^{\pi/2} \xi(r_{\perp}, r_{\parallel})\sqrt{1 - \mu^2}L_{\ell}(\mu){\rm d}\theta,
\label{eq.multipoles}
\end{equation}

\noindent where $\mu = s / r_{\perp} = \cos\theta, s = \sqrt{r_{\perp}^2 +
r_{\parallel}^2}$, and $\theta$ is the angle of $s$ with respect to the
line-of-sight direction. We note that, in linear theory, the redshift-space 
clustering is fully characterized by the monopole ($\ell = 0$), quadrupole ($\ell = 2$)
and hexadecapole ($\ell = 4$).

\section{The properties of LRG and ELG galaxy samples}

\begin{table}
\caption{Global properties of the galaxy samples studied}
\label{table.samples}
\begin{tabular}{@{}lccc}
\hline
Galaxy sample property								& \msg\ 	& \sfg\  \\
\hline
min. stellar mass, \ $M_{\rm stellar}^{\rm cut} \mhalo$ 			&$3.2\times 10^{10} $ & $1.4\times 10^{9}$ \\
min. star-formation rate, \ ${\rm SFR^{\rm cut}} M_{\rm \odot}/{\rm yr}$ 	&$0.0$ & $9.1$ \\
mean stellar mass, \ $\langle M_{\rm stellar} \rangle \mhalo$ 			&$4.5\times10^{10}$ & $2.3\times 10^{10}$ \\
mean star-formation rate, \ $\langle{\rm SFR} \rangle M_{\rm \odot}/{\rm yr}$	&$4.22$ & $14$\\ 
mean halo mass, \  $\langle {M}_{\rm halo} \rangle \ M_{\odot}/h$ 		&$4.8\times 10^{12}$ & $7.8\times 10^{11}$ \\
total satellite fraction, \  $f_{sat}$ 						&$0.22$ & $0.12$ \\
clustering bias, \ $b$ 								&1.84 & 1.02 \\
\hline
\end{tabular}
\end{table}

In this section we explore the overall properties of our galaxy samples.
We start by presenting differences in the distribution of host halo
masses for \msg s and \sfg s (\S3.1) and the radial distribution of satellites
inside these halos (\S3.2). We then show how these differences propagate to the
real-space correlation function (\S3.3) and to the multipoles of
the redshift-space correlation function (\S3.4)

\subsection{The halo occupation distributions}

\begin{figure}
\includegraphics[width=8cm]{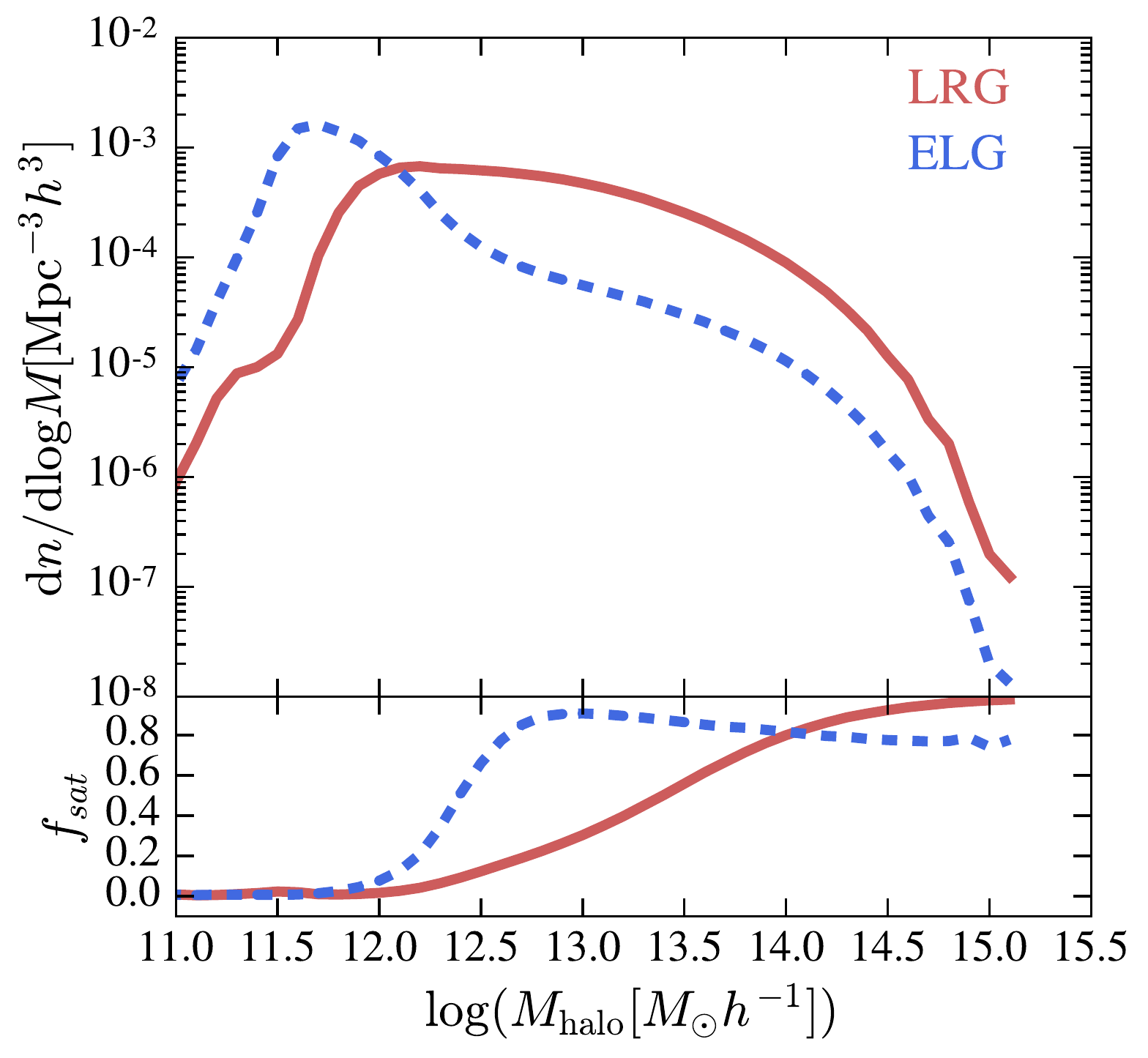}
\caption{Top: the halo mass function of different galaxy samples. \msg\
selection is shown in solid red lines, \sfg\ in dashed blue lines. (See section \ref{sec.gsample} for
details of the construction of each sample). Bottom: The fraction of
satellites as a function of halo mass.}
\label{fig.nm}
\end{figure}

The two galaxy samples described before populate dark matter halos of
different masses and properties. Fig.~\ref{fig.nm} shows the halo mass
distribution of the \msg\ and \sfg\ galaxy samples. Both samples span roughly
the same halo mass range but the \msg\ sample is typically hosted by more
massive halos than the \sfg\ sample.  The average host halo mass is $\langle
M_{\rm halo} \rangle  = 4.8\times 10^{12} \mhalo$ for the \msg\ sample, and
$\langle M_{\rm halo} \rangle = 7.8\times 10^{11} \mhalo$ for the \sfg\ sample.
Additionally, the bottom panel of Fig.~\ref{fig.nm} shows the satellite
fraction $f_{sat}$ of both galaxy samples as a function of the host halo mass.
Below $M_{\rm halo} \sim 10^{12} \mhalo$, the \sfg\ sample is almost
exclusively made up of central galaxies, whereas above that threshold almost
all galaxies are satellites, $f_{sat} \sim 1$, with a slight decrease at higher
masses. On the other hand, a much more gradual increase in the satellite
fraction can be seen in the \msg\ sample: the transition from $f_{sat} \sim 0 $
to $1$ occurs over two orders of magnitude in mass. Table~\ref{table.samples}
summarizes the main properties of both galaxy samples. 

The above trends can be understood in terms of the galaxy formation physics
shaping the star formation and build up of stellar mass of central and
satellite galaxies. At first order, the stellar mass of a galaxy reflects
the total amount of baryons it had available throughout its life to transform 
into stars. Thus, the features seen in the \msg\ sample reflect those of
a population of dark-matter mass selected halos and satellite subhalos: as
we consider halos of higher masses, their abundance decreases smoothly while 
their number of satellites increases. 

On the other hand, the star formation of a galaxy is modulated by feedback and
environmental effects. Massive galaxies in large halos are expected to be more
frequently under strong star-formation regulating mechanisms (e.g. AGN
feedback) and thus, present low star formation rates. Additionally, low star
formation rates are also expected in galaxies hosted by low mass halos due to a
combination of other quenching mechanisms (e.g. supernova feedback) and the
smaller amounts of baryons available. For these reasons most of central
galaxies are located in halos of $\sim 10^{12} \mhalo$. Satellite galaxies, on
the other hand, will be additionally quenched by tidal and ram-pressure
stripping of their hot gaseous halo. The timescale for these is relatively
short, nevertheless, while still active, these satellite galaxies dominate the
number counts over the more massive but low-star forming centrals, which drives
the satellite fraction close to unity above masses $\sim 10^{13} \mhalo$.

%


\subsection{The number density profiles of satellite galaxies}

\begin{figure}
 \centering
 \includegraphics[width=8cm]{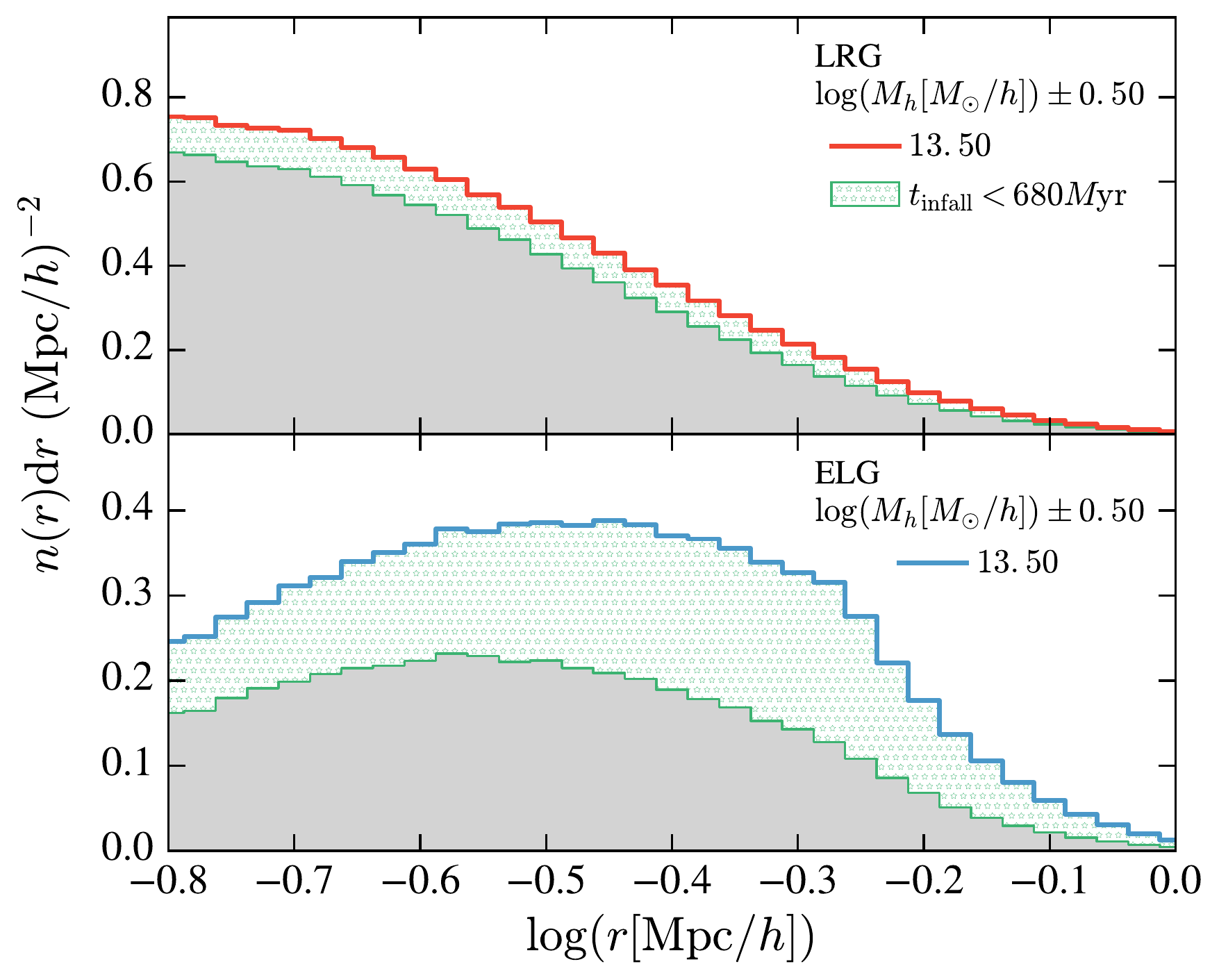}
 \caption{The intra-halo radial distribution of the \msg\ sample (top) and the
\sfg\ sample (bottom) satellites in host halos of
mass $\log M_{\rm halo} [M_{\odot}/h] = 13.5 \pm 0.5$. The green dotted
region indicates the contribution of satellites accreted into the
host halo within the last $\sim 680$ Myr.} \label{fig.rdist}
\end{figure}

The physical processes discussed above not only affect the host halo mass
distribution, but also the regions in the halo preferentially occupied by
different galaxies. As an example, Fig.~\ref{fig.rdist} shows the radial distribution of
satellites within host halos of mass $\log M_{\rm halo}
[M_{\odot}/h] = 13.5 \pm 0.5$. The \msg\ sample features a high density 
of satellites towards smaller radii,  whereas the \sfg\ sample 
peaks at the outskirts of their host halo, $r\sim 0.5 \mpc$. 
Although not shown here, this qualitative picture is found on halos of all
masses.

The origin of such difference is the same physics discussed before: satellite
galaxies can present high star formation rates only for a short period after
the relatively fast-acting tidal and ram pressure stripping has starved the
galaxy of its gas reservoir. Therefore, satellite \sfg\ will be preferentially
located in the outskirts of the halo where recently accreted subhalos are
found. The stellar mass, on the other hand, is more resilient to these effects
and a galaxy can pass our \msg\ selection threshold even after several dynamical
times after accretion.


To show this explicitly, we split satellites according to the infall time to 
their host halo. Those sub-halos accreted within the past $\sim 680 {\rm M}yr$ (green shaded regions
in Fig.~\ref{fig.rdist}), have not experienced their first pericenter passage 
and make up for a significant fraction of the \sfg\ sample. Naturally, these
objects are preferentially found at the outskirts of their host halo.  On the
other hand, galaxies accreted recently are less common in the \msg\ sample and
can be located also in the central regions of the halo. 

We stress that we expect our results to be qualitatively correct regardless of 
the assumptions and parameters of our semi-analytic galaxy formation model. Quantitatively,
however, the lack of star forming galaxies in the center of halos is determined by 
poorly constrained satellite quenching timescales; the cuspiness of \msg s radial
distribution is given by how efficient tidal stripping of stars. This is only approximately
treated in our model after the stripping of dark matter in subhalos. Nevertheless, we anticipate
that all the relevant ingredients will be modelled and constrained better in the future
thanks to observational estimates of quenching timescales \citep[e.g.][]{wetzel10}, 
and comparing to more accurate hydrodynamical simulations \citep[e.g.][]{springel17}.





\subsection{Assembly bias in the galaxy samples}

\begin{figure}
 \centering
 \includegraphics[width=8cm]{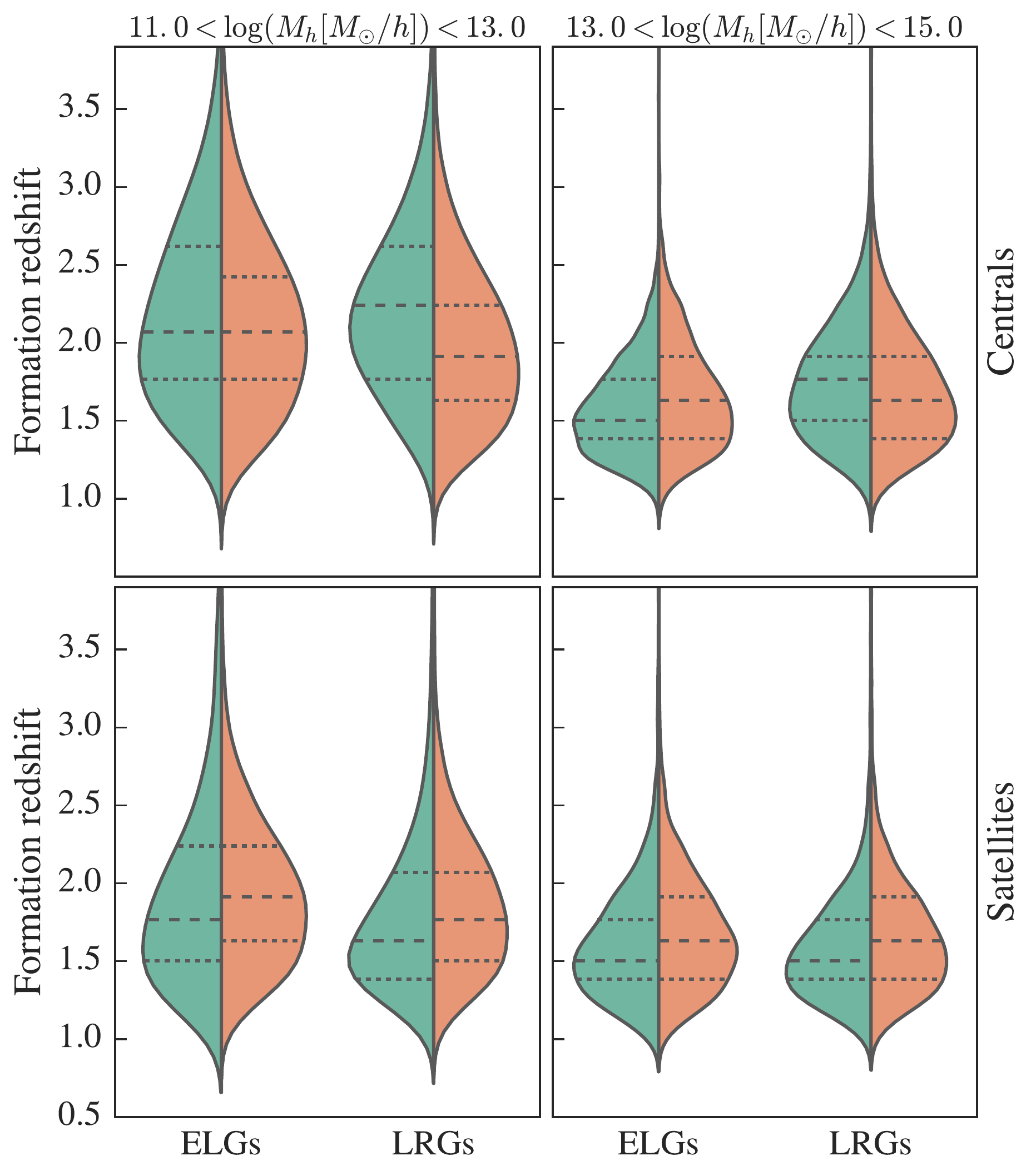}
 \caption{The distribution of formation redshift of the host halos of central galaxies (top)
 and satellites (bottom), for galaxies hosted by halos in different mass ranges, as labeled. 
 Each panel consists of a pair of histograms for the \sfg\ and \msg\ distributions, as shown by the horizontal labels. 
 The formation redshift of host halos for the \sfg\ and \msg\ samples are shown in the green histograms. The orange distributions show the formation
 redshift of a random sample of galaxies, of each corresponding type, drawn from the same host halo mass 
 distribution of the \sfg\ and \msg\ samples, respectively. Horizontal dashed and dotted lines 
 show the median and 25-75 percentiles of the distributions, respectively. } 
 \label{fig.violin}
\end{figure}

Halo clustering is not only a function of the mass, but it also depends on other properties such
as formation time, concentration, spin \citep[see,
e.g.][]{gao05, wechsler06, gao07, angulo08b, lacerna11, hearin16, zehavi17} --
an effect usually referred to as assembly bias. The physical processes discussed before can preferentially select for host
halos with certain properties besides their mass. Therefore, our galaxy 
samples could display clustering statistics that amplify or suppress
assembly bias. 

Fig.~\ref{fig.violin} shows the distribution of formation redshifts (defined as the redshift 
when half of the mass was first acquired) of the host halos of central and satellite galaxies
in our samples. We show results for two host halo mass ranges and for
\sfg s and \msg s, as indicated by the legend. Each of our measured distributions
(green histograms) is compared with a control sample computed using a random 
set with the exact same distribution of halo masses
(orange histograms). 

Central \msg s are preferentially in halos older than the average. This can be a 
consequence of the higher-than-average stellar mass of central galaxies in older 
halos owing to the longer times those galaxies have had to build their stellar content.
This effect is less pronounced at high halo masses, reflecting the fact that almost
all of those high mass halos will contain galaxies that pass our selection
criteria. Central \sfg s in low mass halos have very similar formation redshifts 
as those of the control sample, suggesting that the quenching of central galaxies 
at a fixed halo mass is a somewhat stochastic process. At high masses, on the
other hand, \sfg s are preferentially in younger halos, which is expected because
of the connection between mass accretion and star formation rates.

Satellite galaxies of both types, and in host halos mass ranges explored,
are preferentially found on halos more recently formed than those in the control samples.
Additionally, there does not appear to be significant differences between the
formation times selected by our \sfg\ and \msg\ criteria. Therefore, the
preference appears to be a a consequence of younger halos having a more
abundant satellite population reflecting a larger fraction of their mass to be
recently acquired through mergers \citep[e.g.][]{ zentner05, chaves-montero16,
zehavi17, contreras17}. 

Because of lower satellite fractions and a weaker correlation between formation 
time and star formation rate in central galaxies, we expect {\it assembly bias 
to be less important for \sfg s than for \msg s}. A detailed quantification and
characterization of this effect is beyond the scope of this work. Nevertheless,
the effects discussed above illustrate how galaxy formation couples to halo
properties and is expected to leave imprints in the galaxy clustering on all scales.
Thus, these effects should be incorporated for an accurate and complete modelling 
of galaxy clustering.

\subsection{The real-space clustering of galaxies}

\begin{figure}
 \centering
 \includegraphics[width=8cm]{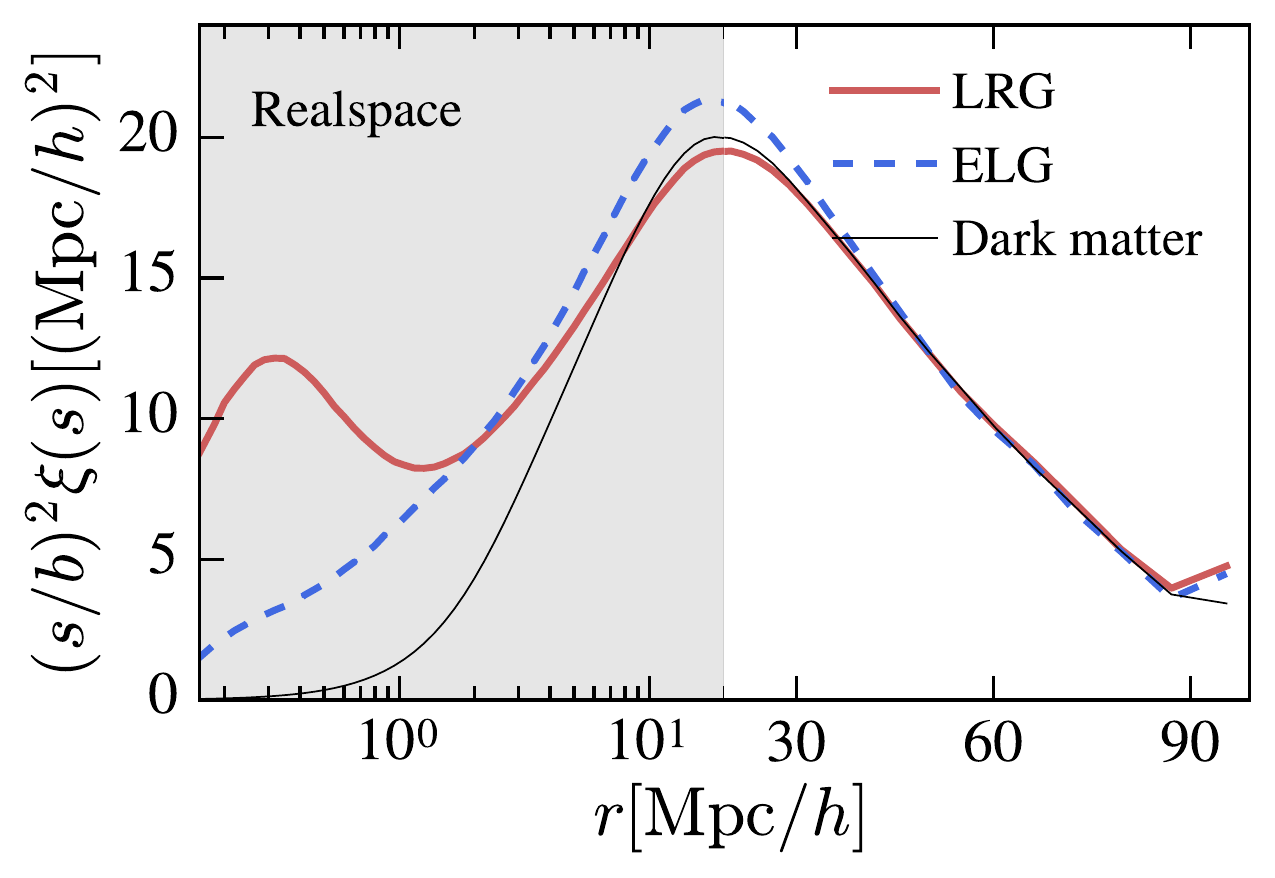}
 \caption{The correlation function of galaxies of the \msg\ (solid red) and \sfg\ (dashed blue)
galaxy samples, divided by $(b)^2$. The
black curve corresponds to the linear-theory correlation function of the dark
matter. The region below
$20 \mpc$ (shaded in gray) is displayed in logarithmic scale along the x-axis,
whereas the region above $20 \mpc$ is displayed in linear scale.}
\label{fig.xinorm}
\end{figure}

The differences in the distribution of host halo mass, formation times, and radial profiles discussed above
result in differences in galaxy clustering. This is shown in
Fig.\ref{fig.xinorm}, where the correlation function of both galaxy samples is
displayed normalized by their respective linear bias parameter squared (making
both curves match at large scales), and multiplied by the scale $s^2$ to
enhance features. Operationally, we compute the bias parameter by taking
the average of the ratio between the real-space correlation functions of each
galaxy sample over the correlation function of the dark matter at scales $40<r
\ [\mpc] < 70$. 

The \msg\ sample features a higher bias, $b = 1.86$, 
than the \sfg\ sample, $b = 1.04$, owing to the higher abundance of massive
host halos compared (c.f. Fig. \ref{fig.nm}). On large scales, $r > 40~\mpc$, both galaxy samples are a linearly
scaled version of the linear dark matter correlation function. On intermediate
scales, however, differences are clearly visible owing to nonlinear galaxy
biases with \sfg\ showing the larger deviations. On small scales,
$r < 10\mpc$ the differences are enhanced due to the different typical host
halo masses and radial distributions: the larger host halos and steeper number
density profiles of \msg\ produce a correlation function with a much more
pronounced 1-halo term.  


%

\begin{figure*}
\centering
\includegraphics[width=17cm]{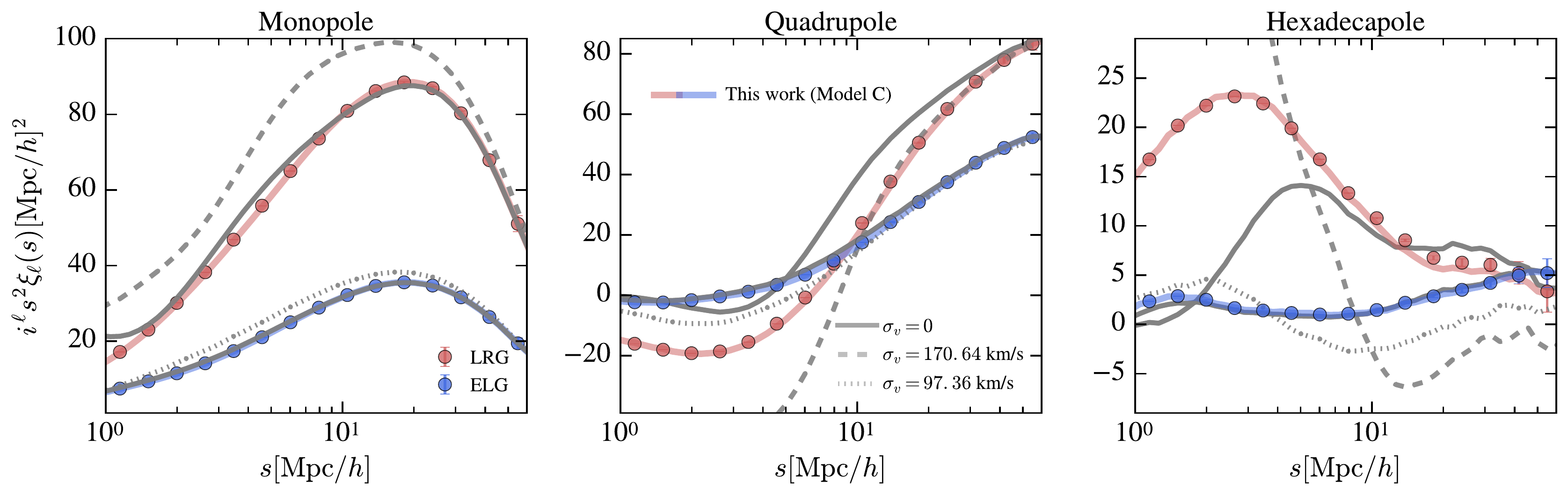}
\caption{The monopole (left), quadrupole (middle), and hexadecapole (right) of
the redshift-space correlation function for \msg s (red circles) and \sfg\ (blue circles).
Gray solid, dashed and dotted lines show the
clustering of the samples assuming no intra-halo velocities, $\sigma_v = 0$;
and assuming $\sigma_v = 170$ or $97 {\rm
km/s}$, respectively. Coloured solid lines display the results of Model C for 
the line-of-sight intra-halo velocity (see \S4 for details).}
\label{fig.xi_s0}
\end{figure*}

\subsection{The redshift-space clustering of galaxies}

The clustering in redshift space is affected by the properties discussed above and
also by the peculiar velocity of galaxies along the line of sight. We show this in
Fig.~\ref{fig.xi_s0}, which compares the monopole ($\ell = 0$), quadrupole $(\ell = 2)$, and hexadecapole ($\ell =4$) of the redshift-space correlation function of our two galaxy
samples. Note that we display the correlation function times $i^2\, s^2$ to
enhance the dynamic range shown.


The differences in satellite fractions and host halo mass can explain part of
the differences we see in Fig.~\ref{fig.xi_s0}: the higher bias of the \msg\
sample implies monopole and quadrupole with larger amplitude compared to 
those of the \sfg\ samples. The differences on smaller scales and for $\ell \neq 0$
multipoles, however, are mostly caused by the differences in the satellite
kinematics, as we will show next. 


\subsubsection{The importance of satellite kinematics}

To quantify the role of satellite kinematics for RSDs, we will compare
the clustering of samples with different distributions of satellite velocities
relative to that of the host halo (thereafter ``intra-halo'' velocities but that
are identical otherwise.

We start by computing a correlation function, $\xi_{\Delta v=0}$,  where all intra-halo 
velocities were set to zero. In redshift space,  by construction this sample of objects displays 
no Fingers-of-God effect, since all velocities are coherent inside a halo. 
We
then convolve $\xi_{\Delta v=0}$ with different intra-halo velocity distribution functions, $\mathcal{P}(\vlos)$:


\begin{eqnarray}
 \xi_{\sigma_v}(r_{\perp},r_{\parallel}) = \int_{-\infty}^{\infty}{\rm d}\vlos \, \xi_{\Delta v=0}\left(r_{\perp}, r_{\parallel} -\frac{\vlos (1+z)}{aH(z)}\right)\,\mathcal{P}(\vlos),
 \label{eq.sigma_conv}
\end{eqnarray}

\noindent and compute the respective multipoles following Eq.~\ref{eq.multipoles}. 

We consider three different $P(\vlos)$:

i) The first one corresponds to a model with zero intra-halo velocity, i.e. $\mathcal{P}(\vlos) = \delta_D(0)$.

ii) The second one describes intra-halo velocities as a Gaussian variate, i.e. 
$\mathcal{P}(\vlos) = \mathcal{G}(0, \sigma_v)$. 

iii) The third one follows a model that will be presented in forthcoming sections 
and that it captures non-zero net infall velocities, host halo mass dependencies, and anisotropies.

To illustrate the impact of intra-halo velocities, in Fig.~\ref{fig.xi_s0} we compare 
the clustering of these three cases (lines) with
the multipoles measured in our original catalogues (symbols). We can see that the case 
with no intra-halo velocities is a fairly good description of the
ELGs multipoles but performs poorly for the LRG sample. This is because the
larger satellite fraction and higher halo mass typical of LRGs compared
to ELGs imply higher typical intra-halo velocities.

The second case, where we set $\sigma_v$ to the standard deviation measured in our catalogues ($170$ and $97 \ {\rm km/s}$ for the LRG and ELG samples, respectively), produces correlation functions that
differ strongly from the true ones. We highlight that modelling small-scale velocities
as Gaussian deviates is a widely-spread practice in large-scale structure and RSD
analyses. Our results, however, indicate that a {\it Gaussian is a very poor description of
the true velocity distributions}. In fact, neglecting intra-halo velocities altogether
is a better assumption because most galaxies in a sample will be centrals
and, thus, are expected to be at rest with respect to the host halo. 


As an additional test, we have repeated our analysis using the value of $\sigma_v$ that
best fits the measured multipoles ($\sigma_v \sim 50\ {\rm km/s}$ for the \msg\ sample
and $\sigma_v \sim 0$ for the \sfg\ sample). In this way, the Gaussian model can provide a reasonable
description of the clustering but only on scales above $s \sim 10-20 \mpc$. 
In general, we expect its
performance to depend on the details of the sample: redshift range, selection criteria, galaxy formation physics, etc, as well as on the combination of multipoles and minimum scale used in the fit. The dependence on these details is simply because considering these velocities as random 
Gaussian deviates is not accurate, and neither physically-motivated, as we 
will show later)

The above illustrates that intra-halo velocities imprint significant features in the multipoles of the correlation function, and that a more sophisticated
modelling given the accuracy and importance of RSD measurements is needed. In the upcoming sections 
we will  focus on developing a minimal but physically-motivated description of intrahalo
velocities. The resulting correlation functions of this model are shown by the solid lines 
in Fig.~\ref{fig.xi_s0}, where we can see that it can correctly capture the behavior
in both \sfg\ and \msg\ samples down to $1~\mpc$. Next we will motivate and discuss the ingredients of the model
and show that it is not only accurate but also flexible enough to capture the expected
diversity induced by different galaxy selection criteria.

\section{An improved description of satellite kinematics}

The analysis of the previous section indicated that an accurate model of the 
redshift-space galaxy clustering should incorporate a correct description of the
small-scale velocities. In particular, we showed that modelling these intra-halo
velocities as Gaussian variables with zero mean produces correlation 
functions that differ systematically from the true ones. We now explore the reason
behind this.




\begin{figure}
\centering
\includegraphics[width=8cm]{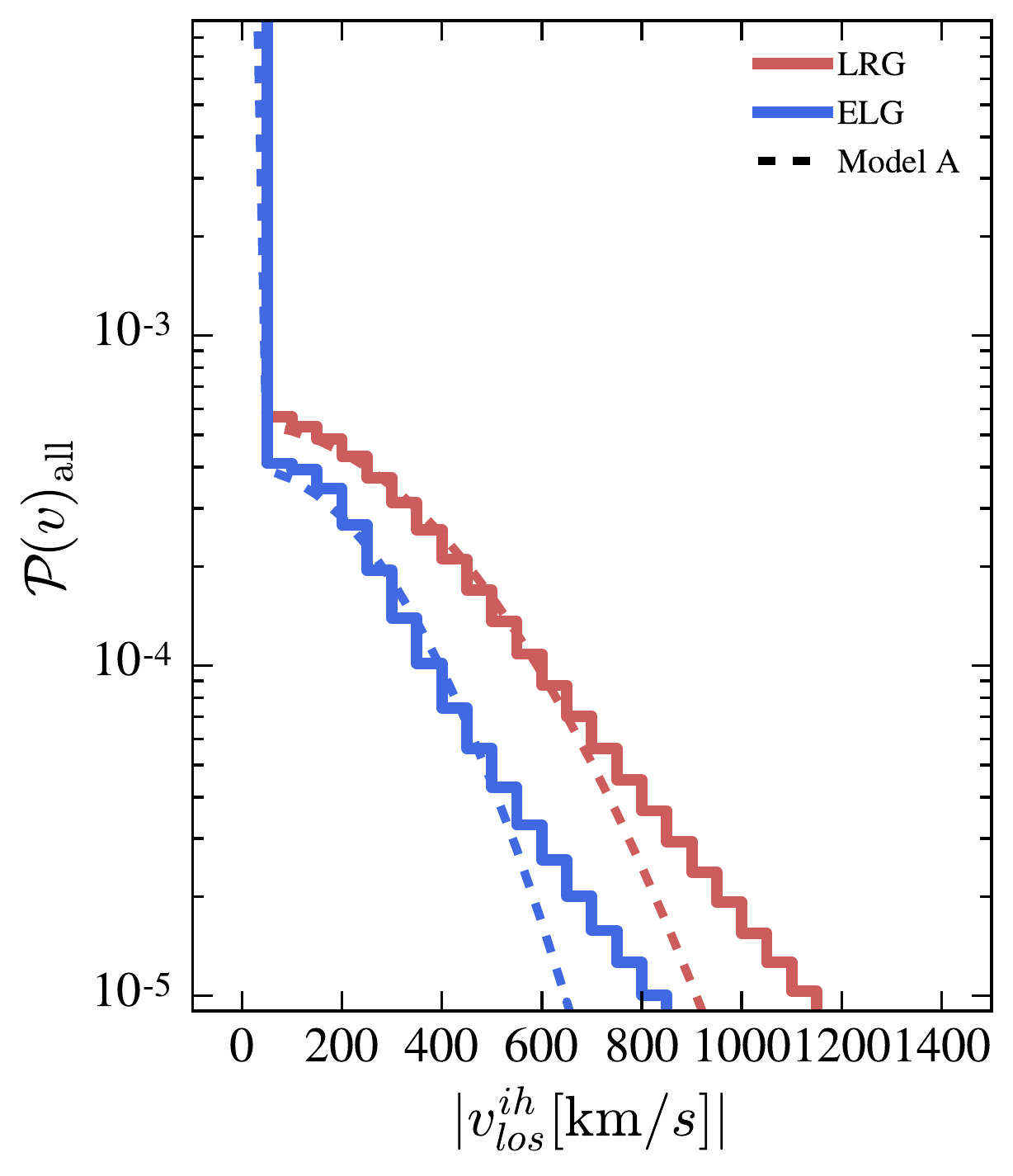}
\caption{The intra-halo velocity distributions of the \msg\ (solid red) and \sfg\ (solid blue) samples.
The dashed lines display Model A, given by Eq.~(\ref{eq.Pvall}).}
\label{fig.vtot}
\end{figure}

In Fig.~\ref{fig.vtot} we show the distribution of the line-of-sight component of 
intra-halo velocities, $P(\vlos)$, for ELGs and LRGs as blue and red histograms,
respectively. Since the fraction of satellites is in both cases small, the
distributions are dominated by a strong peak at zero\footnote{In our model, central galaxies have the velocity of the host subhalo. Note that there are indications that the central galaxy might
not be at rest relative to the host \citep[see, e.g.][]{guo15a}. The expected velocity bias is, however,
small compared to the virial velocity of the host halo, thus we expect this to only 
slightly broaden the distribution of $\vlos$}. Therefore, the term $\sigma_v$ in Eq.
(\ref{eq.sigma_conv}) cannot be directly related to the actual intra-halo
velocity dispersion but to a mixture of velocity dispersions weighted by satellite fractions. All this explains why the Gaussian model did not agree well with the
measured multipoles and why completely neglecting intra-halo velocities lead to a 
more accurate model (c.f. Fig.~\ref{fig.xi_s0}).

\begin{figure}
\centering
\includegraphics[width=8cm]{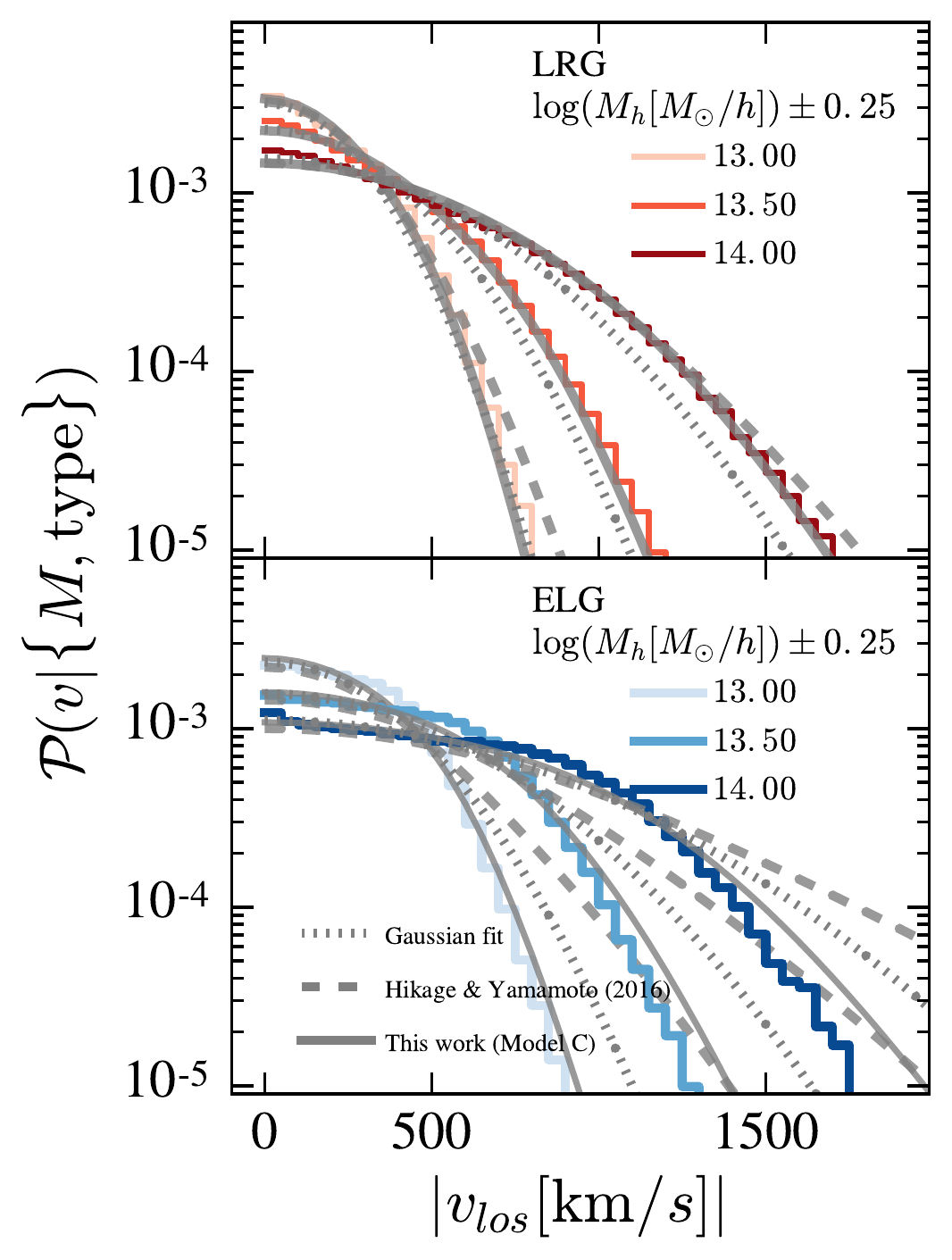}
\caption{The line-of-sight velocity distribution of the \msg\ (top) and the \sfg\ samples (bottom) for satellite galaxies in three host halo mass bins
centered in $\log(M_h) = 13.,13.5.,14.0$, shown by the coloured histograms, with colours according to the legend. The solid gray lines correspond to Model C. The dashed gray corresponds to the model of \citet{hikage16}. The dotted gray line
results from fitting a Gaussian to the distribution of velocities of each galaxy sample.}
\label{fig.vlos}
\end{figure}

\begin{figure}
\includegraphics[width=8cm]{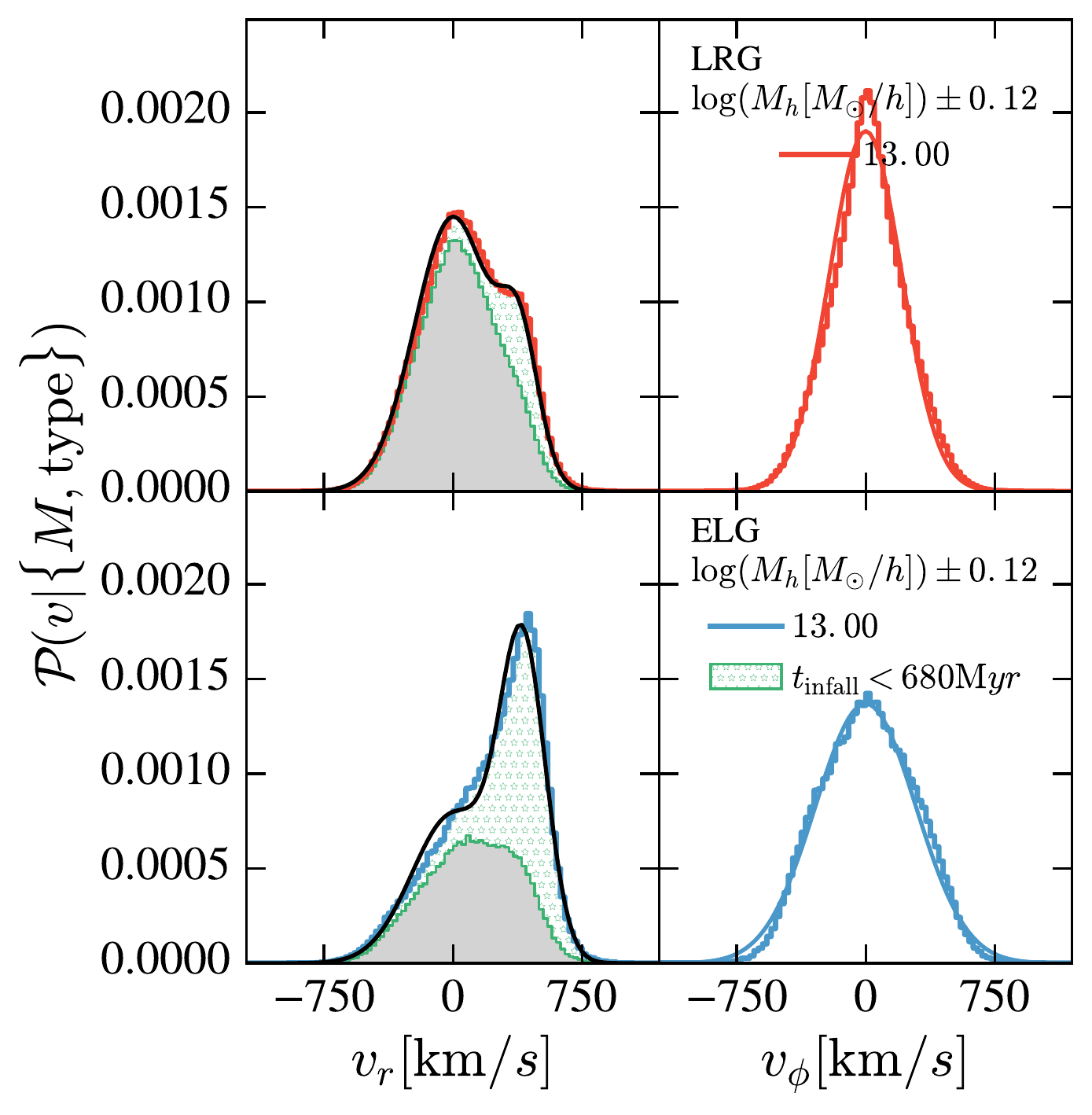}
\caption{The intra-halo velocity distribution components along the radial ($v_{\rm inf}$, left) and the tangential ($v_{\phi}$, right) components. The black solid lines
show the resulting best fit from Model C. The green dotted region corresponds to galaxies accreted within the
last $680 {\rm M}yr$.}
\label{fig.vcomp}
\end{figure}

A simple extension of the Gaussian model is to describe the distribution of intra-halo
velocities as a mixture of a Dirac delta function (representing central galaxies)
and a Gaussian (capturing satellite kinematics): 

\begin{equation}
 \mathcal{P}(\vlos) = (1-f_{\rm sat})\delta_D(0) + f_{\rm sat}\,\mathcal{G}(\mu=0,\sigma),
 \label{eq.Pvall}
\end{equation}

\noindent where $f_{\rm sat}$ is the satellite fraction, $\delta_D$ is the Dirac's delta,
$\mathcal{G}(\mu,\sigma)$ represents a Gaussian
distribution of mean $\mu$ and standard deviation $\sigma$. The dashed lines in
Fig.~\ref{fig.vtot}
show that this model reasonably matches most
of the velocity distribution with only two free parameters ($f_{\rm
sat}$ and $\sigma$). In the following, we refer to this description of
intra-halo velocities as ``Model A''. 

\begin{figure*}
\centering
\includegraphics[width=15cm]{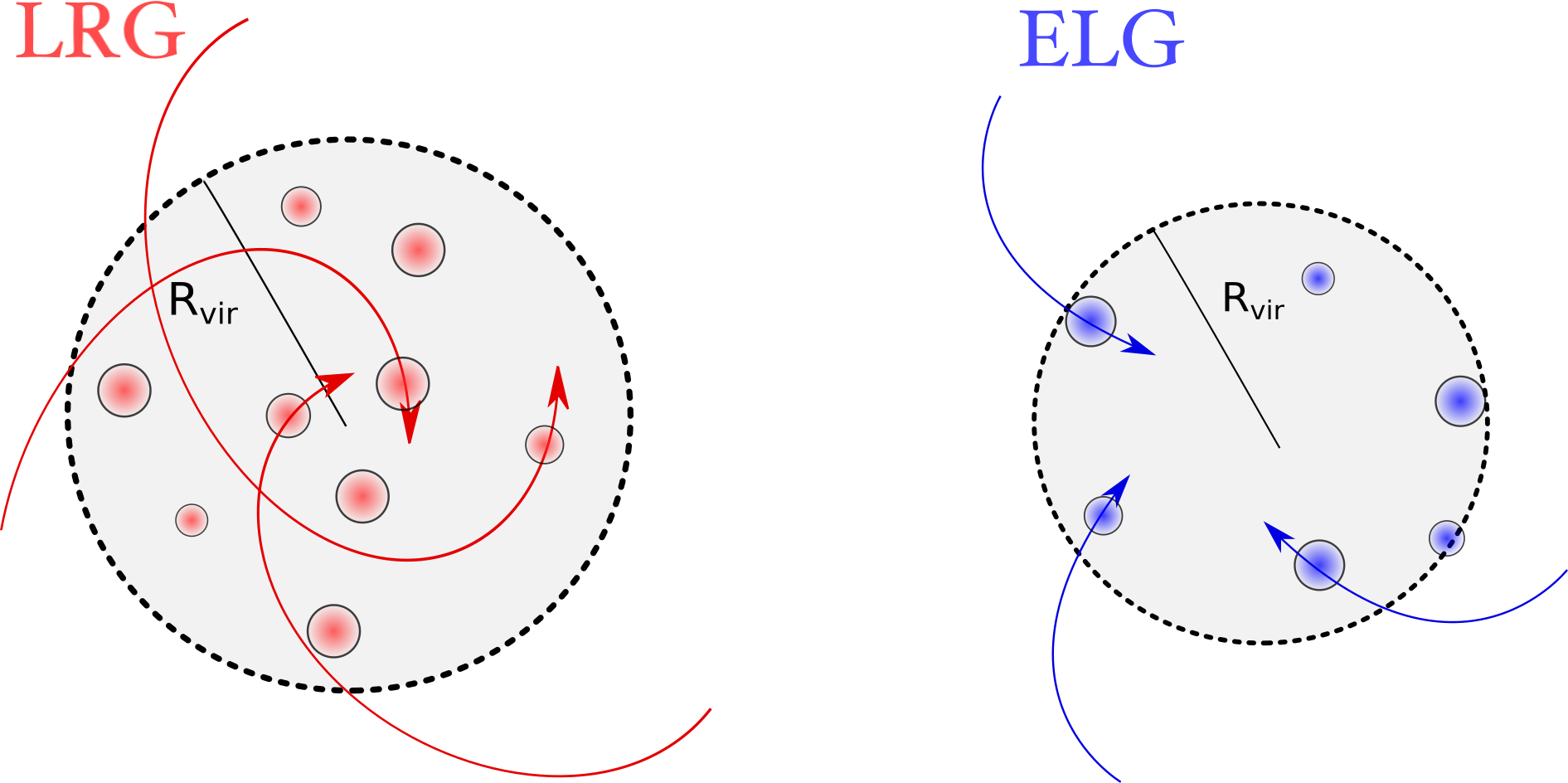}
\rotatebox{0}{\textcolor{red}{\LARGE $\langle M_{\rm halo}\rangle  = 5\times 10^{12} \mhalo$  } \ \ \ \ \ \ \ \vspace{55cm} \textcolor{blue}{\LARGE \hspace{5.25cm} $\langle M_{\rm halo}\rangle = 7\times 10^{11} \mhalo$}}

\caption{A schematic diagram summarizing the properties of the satellites in the \msg\ (left) and \sfg\ (right) samples. The host halo 
is depicted as a gray circle, with the region encompassed within the virial radius $R_{vir}$ delimited with a dashed circumference. \msg\ satellites, shown
as small red circles, are typically hosted by more massive halos compared to \sfg\ satellites, displayed as small blue circles. Satellites in the \msg\
sample are also more numerous, and they are distributed across a broad radial range. \sfg\ satellites are, on the other hand, mostly populating 
the outskirts of their host halos. Arrows illustrate
the trajectories of example satellites.}

\label{fig.cartoon}
\end{figure*}

One limitation of this model is that those galaxies with zero velocities are not necessarily placed at the centres of halos, neither it is guaranteed to have only one of those galaxies per halo.
A more accurate model can thus be obtained if both terms in Eq~(\ref{eq.Pvall}) are associated to their respective galaxy type, assigning zero velocities for central galaxies and drawing velocities from $\mathcal{G}$ for satellites. In the next section 
we discuss this improved description, which we refer to as ``Model B''.

On the other hand, we see that $\mathcal{P}(\vlos)$ has a longer tail than expected for a Gaussian. In Fig.~\ref{fig.vlos} we show $\mathcal{P}(\vlos)$ for satellites inside three
disjoint host halo mass bins, and the respective best fit Gaussian is indicated by dotted lines.
We can see 
that satellites are described by Gaussians of different width, as expected
from the correlation between the velocity dispersion and the mass of halos \citep[see, e.g.][]{wu13}. This naturally creates an extended tail, as the one seen in Fig.~\ref{fig.vcomp} which is made up by the mixture of many distribution functions of different widths weighted by the respective number of satellites. 


Although considering the dependence of $\sigma_v$ with host halo mass is an improved description of the global $\mathcal{P}(\vlos)$, we see that within each mass bin there
are significant deviations from a Gaussian shape. To explore this further, we decompose 
the intra-halo velocity $v$ into a radial and a tangential component, $v_{\rm r}$ and $v_{\theta}$, respectively:

\begin{equation}
 v = -v_{\rm r}\,\hat{e}_r + v_{\phi}\hat{e}_{\phi} + v_{\theta}\hat{e}_{\theta}.
\end{equation}

The velocity along the line-of-sight, $\vlos$ is therefore

\begin{equation}
 \vlos = -v_{\rm r}\,\mu + v_{\phi}(1 - \mu^2)^{1/2},
\end{equation}

\noindent where $\mu$ is the cosine of the angle between the line-of-sight and radial
directions, $\mu = \hat{e}_r \cdot \hat{e}_{\rm LOS}$. By construction, $v_{\theta}$ is chosen
to vanish along the line-of-sight direction.

Fig.~\ref{fig.vcomp} shows the distribution of $v_{\rm r}$ and $v_{\phi}$ for satellite
galaxies in host halos of mass $M_{h} = 10^{13} \pm 0.125 \mhalo$. Tangential velocities are consistent with a single distribution centered at zero, with a somewhat narrower peak for the \msg\ sample. In contrast, for both \msg s and \sfg s,  radial velocities appear
to be
the mixture of two distinct distributions: one centered at zero velocity, 
and another centered at $v_{\rm r} \sim 500$ kms/s (i.e. a net velocity towards the
centre of the halo). The latter distribution is barely noticeable in the \msg\ sample,
but dominant for the \sfg\ sample. 

To clarify the origin of the two populations shown in Fig.~\ref{fig.vcomp}, we have highlighted in green the contribution of galaxies that were
accreted over the past 680 Myr. This shows that the non-zero peak of the radial velocity is created 
by a sub-population of recently accreted galaxies. Consistent to what is shown in Fig.~\ref{fig.rdist}, short quenching timescales favor the \sfg\ sample to be made up significantly by recently accreted objects, which will be located preferentially in the outer
parts of the halo and have net infall velocities.

Motivated by the above, we describe the distribution of $v_{\rm r}$ as a mixture of two Gaussians with different mean, $\langle v_{r,1}\rangle, \langle v_{r,2}\rangle$, and standard deviation $\sigma_{r,1}, \sigma_{r,2}$. In addition, we describe the tangential velocity distributions as a singe Gaussian distribution. Both of these forms, shown by 
black lines Fig.~\ref{fig.vcomp}, agree well with the measured distributions.
Under these assumptions, and if the radial and the tangential velocities are independent, 
the distribution of $\vlos$ is:

\begin{equation}
\label{eq.Pvlos}
\mathcal{P}(\vlos, M) = \alpha f_v(\langle v_{r,1}\rangle ,\sigma_{\mu,1}) + (1 -  \alpha) f_v(\langle v_{r,2}\rangle ,\sigma_{\mu,2}),
\end{equation}

\noindent where $\alpha$ is the relative amplitude of the Gaussians used to describe the infall velocities, and $f(\langle v_{r,i}\rangle , \sigma_{r,i})$ corresponds to

\begin{eqnarray}
\label{eq.fvlos}
               \label{eq.vlos}
f_v(\langle v_{r} \rangle,\sigma_{\mu}) & =&  \frac{1}{2} \int_{-1}^{1} \frac{{\rm d}\mu}{(2\pi)^{1/2}\sigma_{\mu}(\mu, M))} \times  \nonumber \\
               & & \exp \left[ - \frac{(v_{LOS} + \mu\langle v_{r} \rangle (M))^2}{2\sigma^2_{\mu}(\mu, M)}\right],
\end{eqnarray}

\indent where 

\begin{equation}
\sigma^2_{\mu}(\mu, M) = \mu^2\sigma_{r}(M) + (1-\mu^2)\sigma^2_{\phi}(M).
\end{equation}

Note that in the case with no net infall velocities and $\sigma_r = \sigma_{\phi}$, the 
description of line of sight velocities is reduced to a single Gaussian.

We plot the results of this model as solid lines in Fig.~\ref{fig.xi_s0} and 
Fig.~\ref{fig.vlos}.
For computing the parameters of our model, we split our galaxy samples in log bins of halo mass with $\Delta \log M = 0.25$, and then fit two Gaussian components to the infall velocity and a single one to the tangential velocity components. An example of the quality of the fits is shown in Fig.~\ref{fig.vcomp}. The model is able 
to reproduce remarkably well the line-of-sight velocity distribution for both galaxy samples, the agreement being slightly better with the \msg\ sample, with the velocities 
in the \sfg\ sample being slightly overestimated at the tail of the distribution. In the following we refer to this description of intra-halo velocities as ``Model C''. 

A particular case of Model C is when only one population is considered but it is allowed
to have a nonzero net radial velocity. This case was considered by \citet{hikage16} to
describe a mass-selected sample of dark matter subhalos. Fig.~\ref{fig.vlos} shows the performance of this description against the line-of-sight velocity distribution of the two galaxy samples studied here (note that we have fitted again the parameters of the model). 
For the \msg\ sample, this parametrization
results in a reasonable fit, slightly overestimating the tails of the distribution. 
However, for the \sfg\ sample the resulting velocity distributions are a poor fit to the ones in our simulation. This again illustrates how galaxy formation physics combined with
different observational selection criteria preferentially selects satellites in certain
regions in phase space, and thus, that it is necessary to develop flexible and physically
motivated models for satellite kinematics.



\begin{table*}
\caption{Summary of model descriptions of the line-of-sight intra-halo velocity distribution displayed in Fig.~\ref{fig.xileg1}}
\label{table.models}
\begin{tabular}{@{}lll}
Name & Equation 	& Description \\	
\hline
Model $\sigma_v$ & Eq.~(\ref{eq.sigma_conv})  & A Gaussian centered at zero with standard deviation $\sigma_v$. \\ 
Model A & Eq.~(\ref{eq.Pvall}) & Fit to the global line-of-sight velocity distribution. \\
Model B & $\mathcal{P}(v|\ \{{\rm type}\}) = \twopartdef { 0 } {\rm central} {{\rm Eq.} (\ref{eq.Pvall})} {\rm satellite}$  & Zero velocities to centrals, a global Gaussian to satellites. \\
Model C &  $\mathcal{P}(v|\ \{{\rm type}, M_h\}) = \twopartdef { 0 } {\rm central} {{\rm Eq.} (\ref{eq.Pvlos})} {\rm satellite}$  & A two-Gaussian fit to radial velocities and one -Gaussian fit to tangential velocities. \\ 
\hline
\end{tabular}
\end{table*}

Fig.~\ref{fig.cartoon} shows a schematic cartoon that summarizes the differences between satellite galaxies in the \msg\ and \sfg\ samples. Namely, both
galaxy satellites are distributed in a different way within their host halos, and they also display different intra-halo velocities. Environmental 
mechanisms that quench the star-formation in satellite galaxies are responsible for \sfg\ satellites populating the outskirts of their
host halo. Also, the same mechanisms, stronger in massive halos, make these galaxies to populate less massive halos compared to galaxies in the \msg\ sample.

The global picture suggested by Fig.~\ref{fig.cartoon} and detailed throughout this section is robust against the specific modelling of the \citet{guo11}
semi-analytical model. Although environmental processes are modelled differently across different galaxy formation models,
and are not robustly constrained by observational data, there is strong observational evidence for galaxy transformations triggered in dense regions \citep[see, e.g.][]{hashimoto98,kauffmann04}. 
Moreover, the description of intra-halo velocities presented in Eq.~(\ref{eq.Pvlos}) is controlled by parameters that can accommodate a wide range of variations from our baseline model, the \citet{guo11} semi-analytical model. 
Therefore, we expect that our 
model for intra-halo velocities to be flexible and robust enough for interpreting 
future observations.

\subsection{The clustering accuracy with different descriptions of intra-halo velocities}

\begin{figure*}
\centering
\includegraphics[width=17cm]{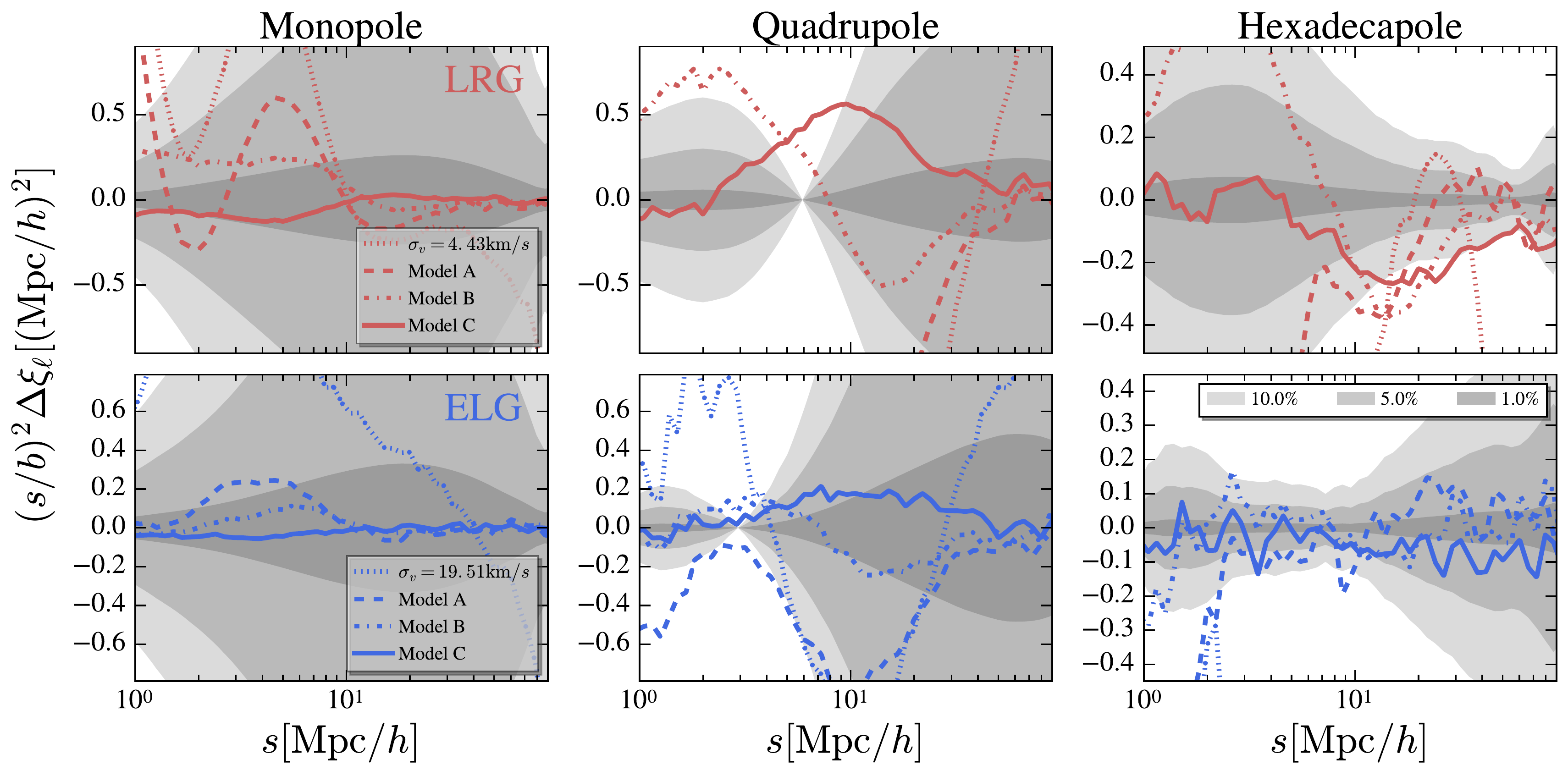}
\caption{The difference between different implementations of intra-halo velocities against the redshift-space monopole 
(left), quadrupole (middle) and hexadecapole (right), as a function of scale. Red and blue lines show the results for the \msg\ and \sfg\ galaxy samples, respectively.
Dashed, dot-dashed and solid lines correspond to Models A, B and C respectively. 
The shaded region in each panel shows where the fractional accuracy is $1, 5$ and $10\%$, according to the legend.}
\label{fig.xileg1}
\end{figure*}

In the previous subsection we explored different models for the intra-halo velocities, which we summarize in Table~\ref{table.models}. We now quantify the accuracy with which they can describe 
the multipoles of the correlation function. Fig.~\ref{fig.xileg1} shows $\Delta \xi_{\ell}$, the difference between the true correlation function and
that predicted by different models for  $\mathcal{P}(\vlos)$, scaled by $(s/b)^2$ to enhance the dynamic range shown. 


First, we test the accuracy of Model A, i.e. a global velocity distribution for each galaxy sample. Operationally, instead of drawing velocities from the functional form of Eq.~(\ref{eq.Pvall}), we shuffle all intra-halo velocities in our catalogues regardless of 
the galaxy type or halo mass, and compute the clustering. 
Overall, Fig~\ref{fig.xileg1} shows that this model is accurate to within 1\% in scales down to $s\approx 10 \mpc$ for the monopole of both galaxy samples. The quadrupole and hexadecapole
are poorly described in scales below $s \approx 20-30~\mpc$. Despite its simplicity, it
already performs substantially better than the case where intra-halo velocities are modelled
as a single Gaussian deviate.

Next we test Model B. We do this by shuffling the intra-halo velocities of all satellites in the galaxy samples, leaving central galaxies unaltered. 
As a result, the clustering improves significantly for both samples, specially for the
quadrupole and hexadecapole. In particular, the accuracy of the monopole increases by
roughly factors of two on scales below 10 Mpc/h.

Finally, we test Model C, which accounts for the halo mass dependence and the infalling velocity component of satellites. 
This description of intra-halo velocities is strikingly accurate within $1\%$ down to scales of 
$s \approx 1 \mpc$ for the monopole of both galaxy samples. The quadrupole of both samples is also accurate except in an intermediate region between $s \sim 3-10 \mpc$
where the accuracy drops above $10\%$. The discontinuity of the shaded regions showing the fractional accuracy around this range in Fig~\ref{fig.xileg1} arises
from the quadrupole value crossing zero in this range. The hexadecapole obtained is 
also fairly accurate, although significantly noisier than the other two multipoles. 

Although not shown here, we have also checked the performance of the special case of simplifying the description of infall velocities with a single Gaussian, as in \citet{hikage16}. This results in
slightly worse multipoles for the \msg\ sample, yet reach a similar accuracy in scales $s \gtrsim 2 \mpc$. However, this model performs significantly worse 
for the \sfg\ sample, reproducing the multipoles only within $5\%$ of of accuracy at scales of $s \approx 2 \mpc$ for the monopole, and within $10 \%$ 
for the quadrupole and hexadecapole for the same scale.


\subsection{Possible improvements}

Although the accuracy of our Model C is already high, we have nevertheless explored
different paths for improving further its performance:

i) The upper limit in accuracy for \modc\ is obtained by employing the exact line-of-sight velocity distribution. We test for this by shuffling the line-of-sight component of the intra-halo velocities of satellites 
among halos of the same mass. However, we have confirmed that this results in only a slight improvement compared to what is obtained with Eq. (\ref{eq.Pvlos}). 

ii) An additional level of sophistication is obtained by incorporating the radial dependence of the intra-halo velocity distribution. 
We have tested this idea by shuffling the line-of-sight velocity distribution of satellites in bins of mass and also three bins
of $r/r_{\rm vir}$. The improvement in clustering accuracy of this model
is negligible compared to what is obtained without considering the radial dependence.

iii) Another possible improvement of our model is to consider correlations with the large-scale velocity field \citep[e.g.][]{guo15a,guo15b}
We have tested this by constructing a sample of galaxies in which the direction of the intra-halo velocity of each galaxy is randomized, thus preserving their amplitude, and any correlation between this velocity and
the super-halo environment. This, however, resulted in only a slight improvement on the \msg\ sample for scales 
of $s \lesssim 2 \mpc$. 


iv) Finally, we explored a description of velocities in which 
the anisotropy of the intra-halo velocity distribution is preserved. 
The anisotropy of the velocity dispersion has been shown to be key to interpret and model
kinematic mass data \citep{wojtak13,kafle14} and can thus lead to further corrections in the 
amplitude of fingers of god, especially if the anisotropy depends on the properties of the galaxy samples. 
To construct these galaxy samples 
we preserve the radial component of the intra-halo velocity of each galaxy, 
and randomize the direction of the tangential component. 
For the \msg\ sample, preserving the anisotropy results in clustering measurements accurate to within $0.1\%$ down to scales of $1 \mpc$. 
The quadrupole and hexadecapole of this model description is consistent to what is found with the catalogue that randomizes the direction of the full velocity vector.
For the \sfg\ sample preserving the anisotropy also results in no significant improvement over \modc\ described by Eq.~(\ref{eq.Pvlos}).

The relatively minor improvements obtained by the above modifications suggests that the 
current version of Model C
is already describing most of the uncorrelated aspects of the intra-halo velocity distributions. A possible direction where more significant gains could be achieved concerns the correlations among the velocity of satellites: For instance, some satellites are
expected to be accreted in groups \citep[e.g.][]{angulo09}, thus their velocity should be highly correlated. We have not attempted to extend our models to accommodate for such given that implementing them in theoretical models of RSDs is not straightforward.

\section{Discussion}
\label{sec.discussion}



In the previous section we have proposed several models for the intra-halo velocity
dispersion of satellite galaxies. All of these models resulted in accurate but different
predictions for the multipoles of the redshift-space correlation function. We now compare 
and quantify the improvements by means of the minimum scale, $s_{\rm min}$, above which 
they would deliver a good fit to the measured correlation functions. In practice, we first
compute  

\begin{equation}
 \chi_{\nu}^2(s) = \frac{1}{3N_s - N_p}\sum_{i,j > s}\left[w_i^{samp} - w_j^{mod}\right] C_{ij}^{-1}(V) \left[w_j^{samp} - w_i^{mod}\right]^T,
\end{equation}

\noindent where $N_s$ corresponds to the number of measurements in each multipole, $N_p$ the number of free parameters of the model, $w_i^{k} = [\xi_0(s_i), \xi_2(s_i),\xi_4(s_i)]$ with $k$ referring
to either a galaxy sample or a model. $C_{ij}^{-1}(V)$ is the inverse of the covariance matrix computed according to Eq.~(\ref{eq.cov}) rescaled to a volume $V$.
Then, we find the value
of $s_{\rm min}$ as $\chi_{\nu}^2(s) < 1$ for all $s > s_{\rm min}$. Therefore, $s_{\rm min}$
represents the smallest scale that can be consider before an hypothetical fit starts delivering
biased constrains. 

\begin{figure}
\centering
\includegraphics[width=8cm]{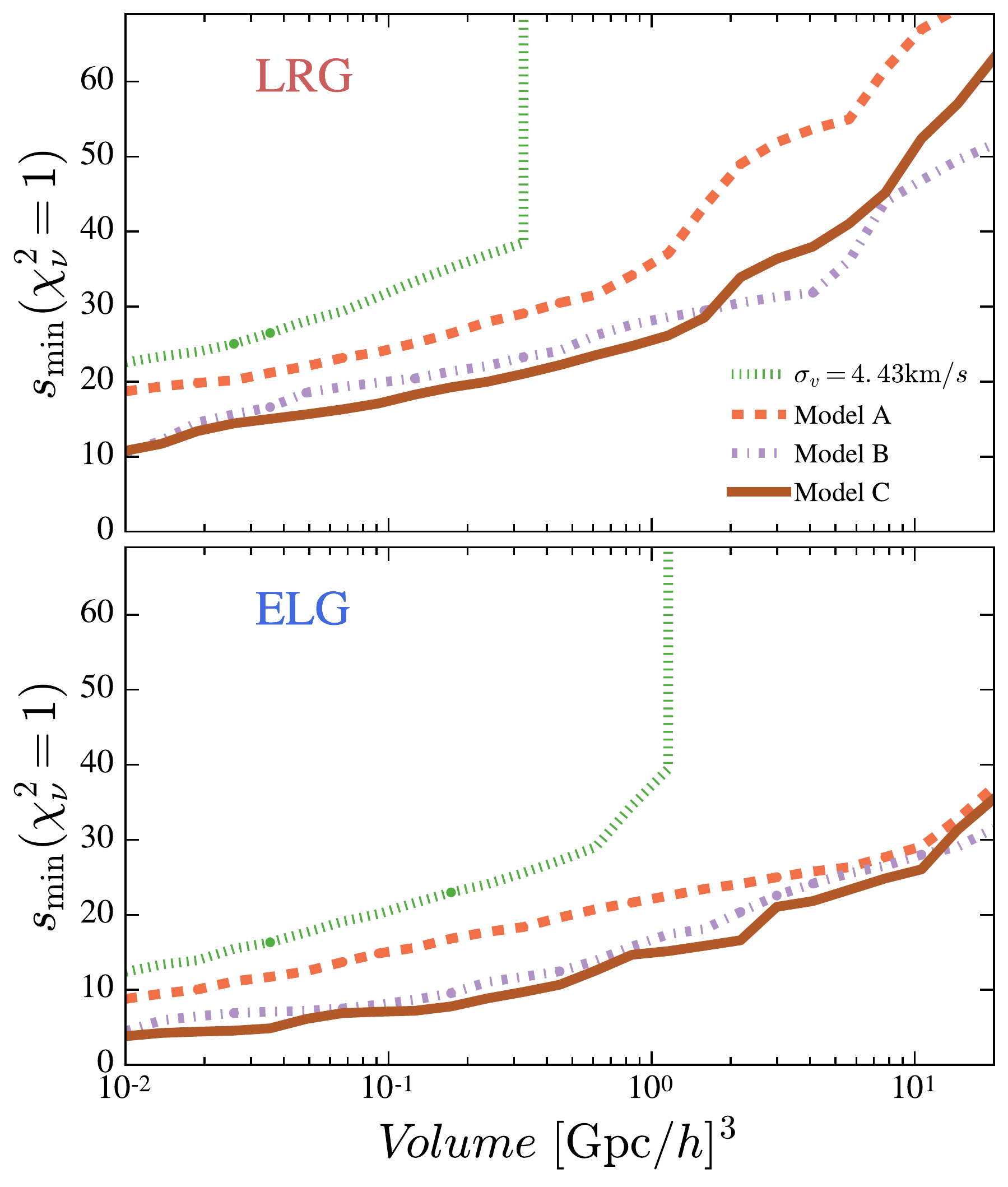}
\caption{The optimal scale to perform a likelihood analysis for which $\chi^2_{\nu} = 1$ for different models as a function
of the volume probed. 
The dotted green line corresponds to a constant $\sigma_v$ value that best fits the redshift-space clustering. The
dashed orange, dotted-dashed purple and solid brown lines correspond to Models A, B and C, respectively. The top and bottom
panels correspond to the \msg\ and \sfg\ samples, respectively.
}
\label{fig.chisq}
\end{figure}

Fig.~\ref{fig.chisq} shows $s_{\rm min}$ computed for different models as a function of the
comoving volumes. Small differences between the model and the measured clustering become 
more important as the volume of the survey increases since the elements of the covariance
matrix decrease, resulting in an increase of the value of $\chi^2_{\nu}$ at a fixed minimum scale $s_{\rm min}$.

For all models, the minimum scale providing a good fit is smaller for the \sfg s than for 
the \msg\ sample. Since the \sfg\ sample contains a smaller fraction of satellites, any inaccuracy on the description of intra-halo velocities has a greater impact on the \msg\ 
sample. Consequently, the models for the \sfg\ sample display overall a higher accuracy, as shown also in Figs.~\ref{fig.xileg1}.


For any given volume, Model C reaches smaller scales with good fits compared to all the other models. This is particularly true for the model of the \sfg\ sample. 
For a survey with $V = 10 ({\rm Gpc/}h)^{3}$, comparable to DESI at $z=1 \pm 0.1$ \citep{weinberg13}, 
these limits become $s_{\rm min} \approx 25, 50\ \mpc$ for the \sfg\ and \msg\ samples, respectively.

For illustration, we compare the performance of our models against Model $\sigma_v$. This shows the common strategy in which intra-halo velocities are described by a single 
Gaussian with standard deviation $\sigma_v$ (Eq.~\ref{fig.xi_s0}). 
The value of $\sigma_v$ is chosen to provide an overall good fit to all multipoles between $s = 5-90\ \mpc$. 
Fig.~\ref{fig.chisq} shows that this model performs
poorly, with values of $s_{\rm min}$ systematically larger by a factor 2x and more compared to Models B and C. Moreover, 
for volumes larger than $V \gtrsim 1 ({\rm Gpc/}h)^{3}$, this model delivers $\chi^2_{\nu} \gg 1$ at all scales. 

The regime of large volumes where {Model $\sigma_v$} breaks down is precisely the one that is expected to be probed by future surveys.
Model constraints with $\chi^2_{\nu} > 1$ represent a sub-optimal exploitation of the measurements. 

In contrast, the values obtained for $s_{\rm min}$ with Model A span $10- 30\ \mpc$ for small volumes, 
but reach to $\approx 70\ \mpc$ for the \msg\ sample for large volumes, $V \sim 10 ({\rm Gpc/}h)^3$. On the other hand, {Model B and C} perform similarly. 
This is explained because at large scales both models deliver similar accuracies, as shown in Fig.~\ref{fig.xileg1}.

As discussed before, even though our models assume the correct satellite fraction, halo
occupation distribution, host halo velocities, and radial distributions, these are still
erroneous on small scales given the huge accuracy of upcoming surveys. Even though it might
be possible to improve the modelling further, a more practical approach would be to characterize
and incorporate the uncertainty of the theoretical models directly in the data analyses.

\begin{figure}
\centering
\includegraphics[width=8cm]{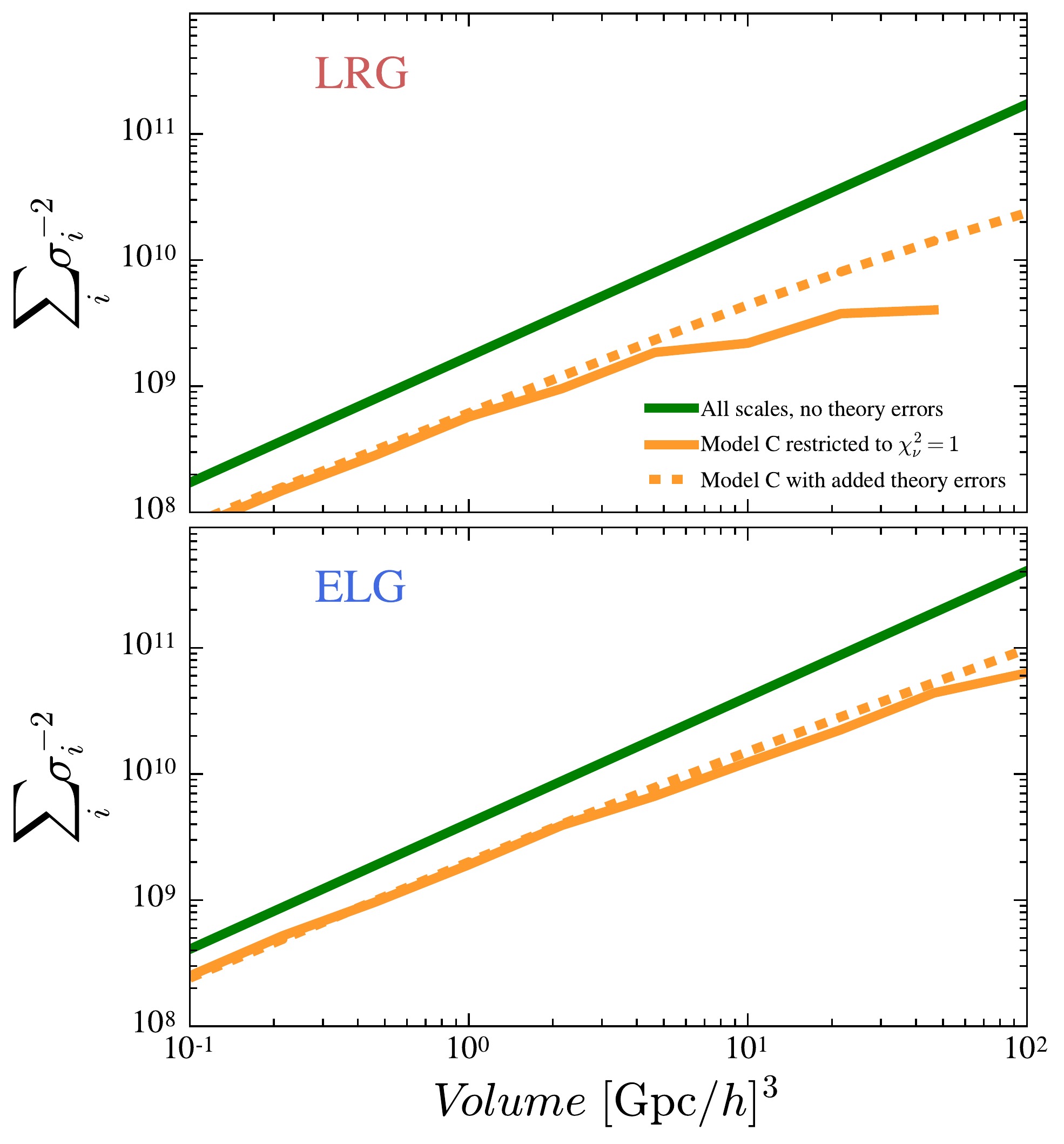}
\caption{The sum of the inverse of the variance terms squared as a function of volume for 
different covariance matrices. The solid green line shows sum over the errors of in scales, taken from the diagonal of the full covariance matrix. The
solid orange line shows the result of restricting the sum to the scales where $\chi^2_{\nu} = 1$ for Model C. The dashed line
shows the result of using all scales adding a theoretical error (see text). Top and bottom panels correspond to
the \msg\ and \sfg\ samples, respectively.}
\label{fig.det_err}
\end{figure}

We explore this idea by quantifying this ``theory error'' as $\sigma_t^2 = 2(\Delta \xi_{\ell})^2$, with $\Delta \xi_{\ell}$ the difference between the clustering of the model and the galaxy samples (shown in Fig.~\ref{fig.xileg1}). We have then added it to the measured covariance matrix and quantified the constraining power as the sum of the inverse of the
diagonal elements of the covariance matrix. We show the results in Fig.~\ref{fig.det_err}.

In the ideal case of having a completely accurate theory, the error budget would be dominated solely by the statistical 
variance of the measurements quantified by the covariance matrix. On the other hand, restricting the scales to those where $\chi^2=1$ results in a loss of information from small scales. Adding the theory errors to the statistical ones allows to include all scales, and the overall error budget
improves with respect to the case where the scales are restricted. This is shown by the dashed line in Fig.~\ref{fig.det_err}. Although this idea is implemented in a rudimentary form, it might be a promising path to be explored in the future, which would allow a more complete
exploitation of future surveys. 


Finally, we would like to emphasize that our results are based on the \citet{guo11} 
semi-analytical model, which represents the state-of-the-art in galaxy formation 
models. This model captures of all the key physical processes that are expected to shape 
the evolution of galaxies, and shows a remarkable agreement with the many observed 
statistical properties of galaxy populations. However, the predictions for the differences between the \msg\ and \sfg\ samples are 
expected to vary in detail across different galaxy formation models \citep[see, e.g.][for a detailed
comparison between models]{knebe15}. Galaxy formation is a complex process which combines astrophysical mechanisms spanning different 
dynamical ranges to account for observed properties. Thus, important ingredients to our analysis, such as the strength of ram-pressure stripping or feedback processes
are not robustly constrained. Moreover, some of the parameters controlling different physical mechanisms can be degenerate \citep{bower10,lu11,henriques13,ruiz13}.

Despite this quantitative limitation of the models, our descriptions of intra-halo velocities are general enough to accommodate the relevant  
features of the galaxy sample targeted. For instance, regardless of the accuracy of the 
gas stripping prescription in \citet{guo11}, an \sfg\ galaxy selection is likely to result in a population of recently infalling
satellites in haloes with a dominating infall velocity. Such population should be well described with Eq.~(\ref{eq.Pvlos})
to model the distribution of line-of-sight intra-halo velocities in the real Universe. 

The free parameters of models A, B and C can be constrained jointly with
 cosmological parameters using anisotropic clustering measurements. However, extending the analysis to smaller scales may
 not necessarily result in improved cosmological constraints given the number of free parameters. In particular, for current survey volumes, the gain of models
B and C over A is marginal. However, as Fig.~\ref{fig.chisq} illustrates, the gain becomes substantial 
when volumes exceed $\sim 1({\rm Gpc}/h)^3$ and improvements over models with a single Gaussian become necessary. In addition, current progress on perturbation theory models
and numerical approaches are likely to improve the current accuracy of the real-space clustering towards smaller scales \citep{dodelson16}.

Additional priors to the nuisance parameters of our model descriptions can be obtained with theoretical galaxy formation models or by independent datasets.
For instance, one such technique is discussed in \citet{decarvalho17}. Their analysis of SDSS data shows that groups and clusters with non-Gaussian velocity distributions
typically display a higher rate of infalling galaxies. This is consistent with our findings for the ELG sample. Future analysis over large datasets
targeting ELGs, such as eBOSS and DESI, could thus add important priors on the free parameters of our model descriptions.




\section{conclusions}

The next generation of cosmological galaxy surveys requires a significant improvement of 
theoretical descriptions of galaxy clustering. The exploitation of cosmological surveys can 
currently be considered sub-optimal, meaning that only linear or mildly non-linear scales 
are used to constrain cosmological parameters, and only an approximate treatment of the
galaxy velocities are considered. As a result, small scales that are measured with 
the greatest precision are discarded from the analysis. 

Current efforts towards improving theoretical models have focused mostly on improving the 
description of the density and velocity field for dark matter and halos. 
In this paper we show that such approaches will be limited by the particular features
imprinted by galaxy formation physics. Future experiments will target very different types 
of galaxies to perform cosmological analysis, which will increase the importance of
understanding the impact of galaxy formation. 

In this paper we studied the impact of galaxy selection on satellite kinematics and
redshift-space clustering. We made use of a state-of-the-art semi-analytical galaxy 
formation model \citep{guo11} run over one of the largest N-body simulations to date 
\citep[M-XXL, ][]{angulo12}. We build two galaxy samples at $z=1·$ with a fixed number 
density, $n = 10^{-3} ({\rm Mpc/}h)^{-3}$ and a selection criteria based on stellar 
mass and star-formation rate limited samples. These samples represent proxies for 
\msg s and \sfg s, respectively


These galaxy samples feature different halo mass distributions and satellite fractions 
(Fig.~\ref{fig.nm}). The star-formation rate is typically quenched in satellite galaxies 
by gas-stripping processes such as ram-pressure. Environmental processes thus largely 
reduce the abundance of satellite galaxies in the \sfg\ sample and prevent them from 
populating massive haloes, where these effects have a stronger impact. Their typical 
halo mass is, thus $M_h \approx 7\times 10^{11} \mhalo$. On the other hand, the stellar 
mass is tightly correlated with the parent dark matter halo mass, and it is not
affected strongly by environmental effects. As a result, the \msg\ sample is characterized 
by massive dark matter haloes and a higher satellite fraction.


The clustering in real space of these two galaxy samples can be characterized by their clustering bias. 
At large, linear scales, a constant bias parameter is sufficient to describe their clustering accurately. 
However, at smaller scales, $r \lesssim 20 \mpc$  their clustering differs 
significantly on each galaxy sample, as shown in Fig.~\ref{fig.xinorm}.


In redshift-space, the clustering depends also on the peculiar velocities of galaxies. Our galaxy formation
model predicts that galaxy samples have different peculiar velocities. In particular, 
the distribution of intra-halo velocities of the different galaxy samples has a strong 
impact on the clustering in scales well beyond the 1-halo term.


One of key differences between the \msg\ and \sfg\ samples is a population
of recently accreted satellites infalling into their parent halo. These objects are fairly ubiquitous 
in the \sfg\ sample because they have not been significantly affected by environmental effects 
that quench their star-formation yet. The satellite galaxies in the \sfg\ sample are thus comprised of two populations with
different properties: one corresponds to objects affected by gas stripping processes but 
still forming stars at a rate that allows them to be included in our selection criteria. These objects have infall velocities
well described by a Gaussian centered at zero, and occupy a broad range of radial positions within their parent halo. The other
population corresponds to recently accreted satellites populating the outskirts of their parent halo (see Fig.~\ref{fig.rdist})
with a dominant infalling velocity component (see Fig.~\ref{fig.vcomp}).


We develop a description of intra-halo velocities that can account for two populations with different infall velocity
distributions. Eq.~(\ref{eq.Pvlos}) describes the line-of-sight velocity distribution of satellites as a 
function of mass of their parent halo. 
Fig.~\ref{fig.vlos} shows that this description can successfully reproduce the intra-halo line-of-sight 
velocity distribution of the \msg\ and \sfg\ samples. More importantly, we show that this description results in
a predicted redshift-space clustering that is remarkably accurate. For the monopole of the redshift-space correlation function, 
this description is accurate to within $1\%$ in scales of $s = 1-90~\mpc$.


In contrast, the standard approach to account for the impact of intra-halo velocities in
the redshift-space clustering consists in applying a Gaussian smoothing to the clustering, 
with a velocity dispersion term taken as a nuisance parameter. This approach is shown to 
result in poor fits to small scales. Even when considering a model in which both the positions
and velocities of galaxies are exact, except for intra-halo scales, this description limits
the accuracy to scales $s \gtrsim 20-30~\mpc$. Further improvements to this simplistic model 
include modelling the true distribution of line-of-sight intra-halo velocities, which features
a strong peak at zero velocities corresponding to central galaxies. This extension of the model
can improve the accuracy of the clustering to scales of $s \approx 10~\mpc$ for the monopole. 
Further improvements are obtained by separating galaxies in centrals and satellites, and 
drawing their intra-halo velocities as a function of parent halo mass. More detailed improvements, 
such as preserving the anisotropy of the velocities in haloes do not improve the accuracy of the clustering in redshift space
significantly.

Implementing these ideas is not trivial in the standard framework of models used to interpret 
clustering measurements. For instance, analytical models typically do not distinguish satellites and centrals, and so 
the velocity dispersion term introduced effectively affects the whole galaxy sample used \citep[e.g.][]{reid12}.
A more accurate modelling technique could be achieved by means of constructing quick galaxy samples with an 
halo occupation distribution or sub-halo abundance matching technique that incorporate the ideas described above. In this way, 
the non-linear dynamics would be accurately accounted for by creating galaxy samples from an ensemble of N-body simulations
of varying cosmological parameters or by re-scaling techniques that modify the cosmology in a simulation \citep{angulo10}.
These ideas will be investigated further in a future work.

\section*{Acknowledgments}
We acknowledge useful discussions with Simon White, Carlton Baugh, Sergio Contreras,
Andrew Hearin and Andrey Kravtsov. 
AO and RA acknowledge support from project AYA2015-66211-C2-2 of the Spanish Ministerio de Economia, Industria y Competitividad. 
The Millennium Simulation databases used in this paper and the web application providing online access to them were constructed as part of the activities of the German Astrophysical Virtual Observatory (GAVO).
We thank Gerard Lemson for his help optimizing our queries in the Millennium database.
The numerical analysis of this research used Astropy, a community-developed core Python package for Astronomy \citep{astropy13}, Scipy \citep{scipy}
and made use of Matplotlib \citep{hunter07} for the figures.

\appendix
\section{Validation of our results against the Millennium simulation.}
In section 2.1 we discussed the corrections that were necessary to account for the limited mass resolution of the 
catalogues based on the M-XXL run. Several quantities have been compared and adjusted to retrieve consistent results with 
the same model run in the Millennium simulation.
Such tests are described in \citet{angulo14}.
 
However, a key quantity throughout this paper, the distribution of intra-halo velocities, has not been confronted
against the Millennium simulation before. Hence, we 
compare the intra-halo velocity distributions predicted by the \citet{guo11} model based on the M-XXL simulation to 
those from the Millennium simulation. To access the model predictions of the latter, we make use of the 
Millennium database \citep{lemson06}

\begin{figure}
 \includegraphics[width=8cm]{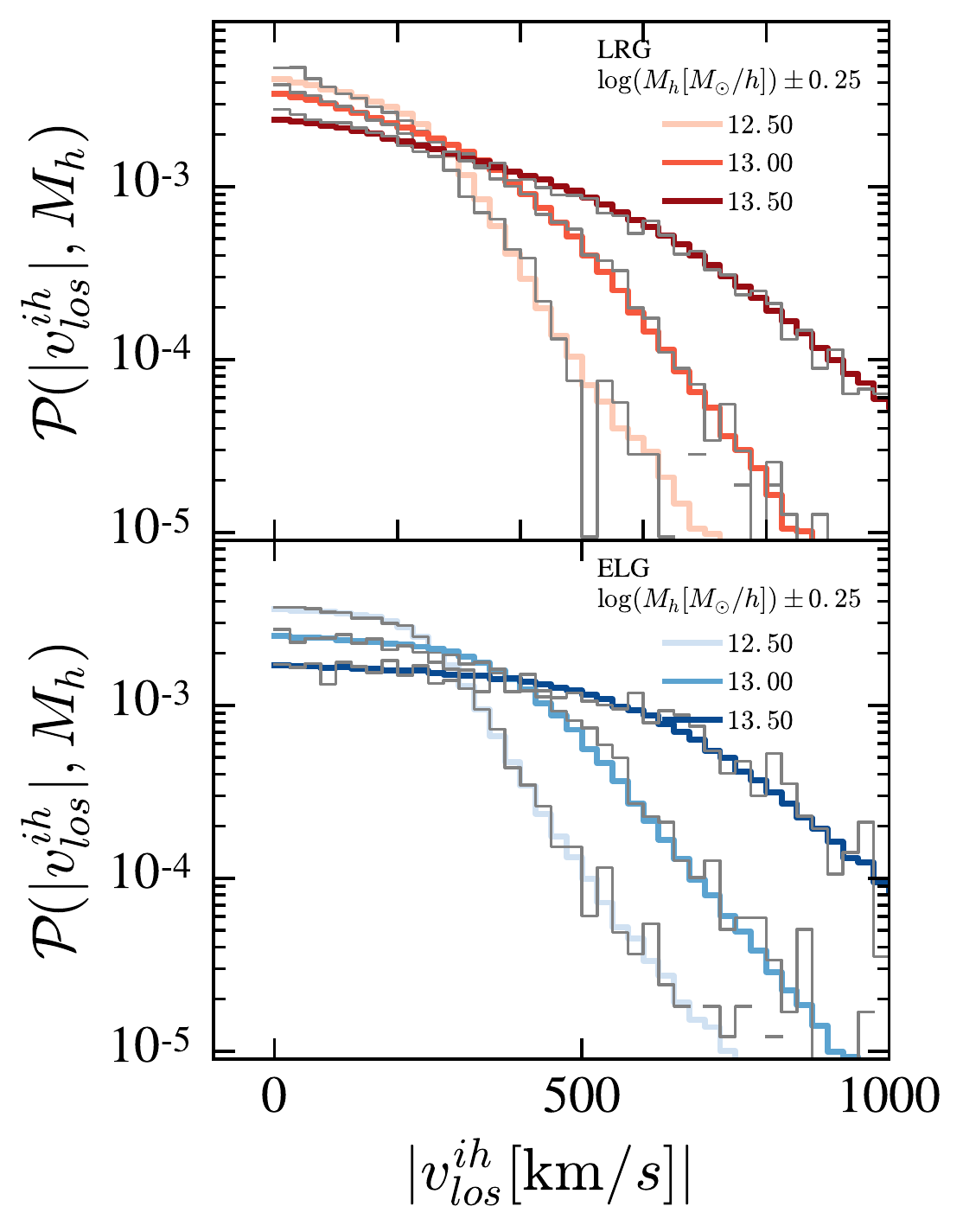}
 \caption{The line of sight velocity distribution of \msg\ (red) and \sfg\ (blue) samples at different halo masses, as shown by the legend.
 The gray histograms show the results obtained using the \citet{guo11} model run over the Millennium simulation.}
 \label{fig.a1}
\end{figure}

Fig. \ref{fig.a1} shows the intra-halo line-of-sight velocity distribution of the two galaxy samples for different host halo masses. 
The same distribution is computed in both simulations. The results between both catalogues match remarkably well.
Evidently, due to the smaller volume, the histograms obtained from the Millennium simulation are more noisy, but there is not any
systematic deviation between both predictions. Therefore, this comparison validates our predictions for the shape of the line-of-sight
velocity distribution against resolution effects.

\begin{figure}
 \includegraphics[width=8cm]{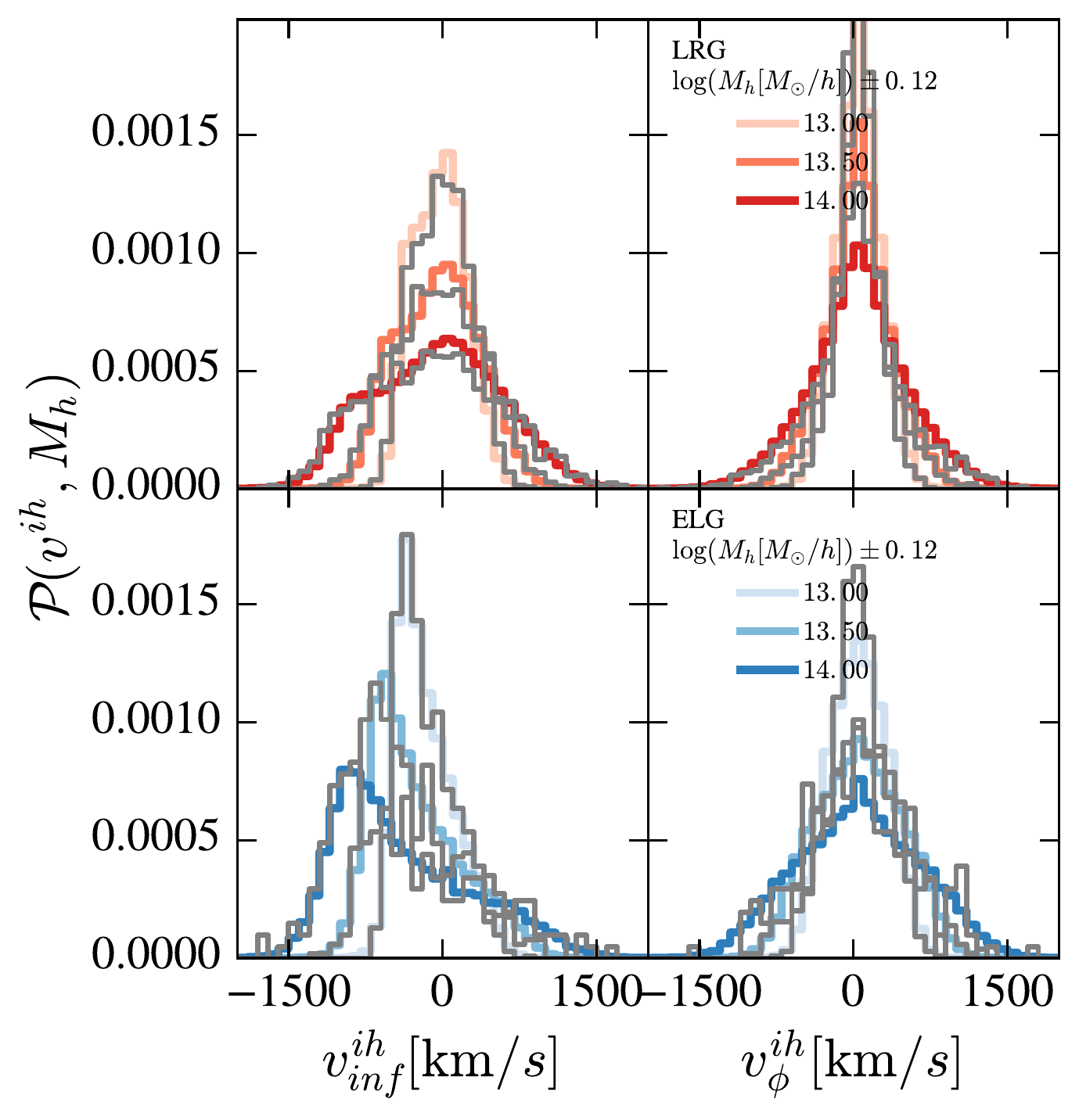}
 \caption{The line of sight velocity distribution of \msg\ (red) and \sfg\ (blue) samples at different halo masses, along the radial (left) and tangential (right) components, similarly to Fig.~\ref{fig.vcomp}.
 The gray histograms show the results obtained using the \citet{guo11} model run over the Millennium simulation.}
 \label{fig.a2}
\end{figure}

To explore the predicted intra-halo velocities in more detail, Fig.~\ref{fig.a2} shows the velocity distribution 
splitted into their infall and tangential components. Due to its large size, particles in the M-XXL simulation were not stored. Thus, there is no information in the simulation to 
assign velocities to the galaxies once they become type 2 (orphans). This creates an artificial peak of positive infall velocities not related
with the population of recently accreted satellites, as discussed in Fig.~\ref{fig.vcomp}, but instead with the moment in which those subhalos were disrupted in the 
simulation to an extent that they could not be identified anymore. Randomizing the direction of the velocity vector results in a remarkable match
of these two velocity components with what is obtained with the Millennium simulation.

In summary, both tests described above show that the velocity distributions and the results over which this paper is based on are not sensitive to 
the limited resolution of the M-XXL simulation.

\begin{figure}
 \includegraphics[width=8cm]{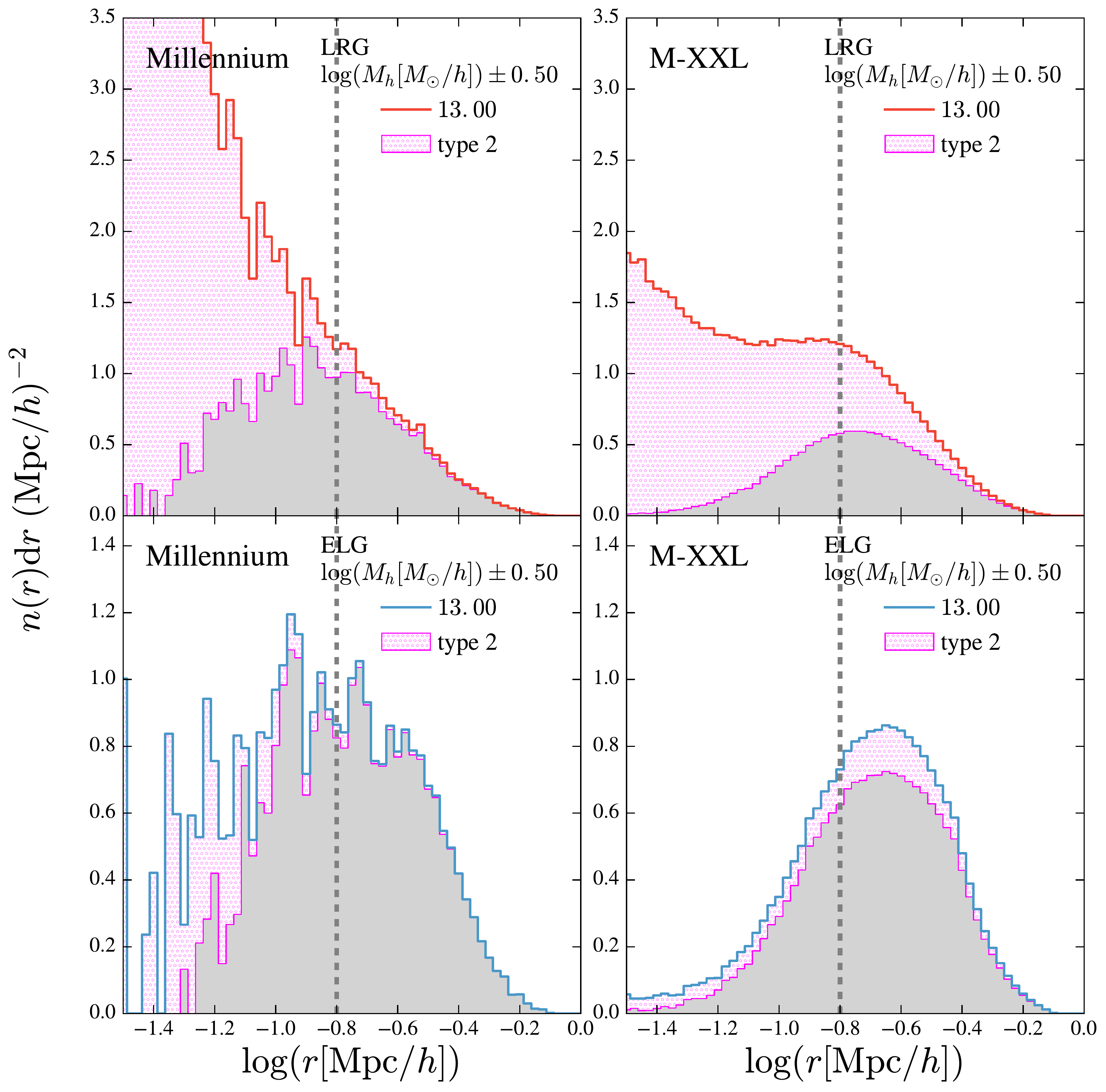}
 \caption{The radial distribution of \msg\ (red) and \sfg\ (blue) samples, similarly to Fig.~\ref{fig.rdist}. 
 The left panels show the results obtained using the \citet{guo11} model run over the Millennium simulation, 
 and the right panels over the MXXL simulation (this work). The pink dotted region corresponds to type 2 satellites.}
 \label{fig.a3}
\end{figure}

Finally, the lack of information in the simulation to assign positions for type 2 galaxies also arises in the overall 
radial distribution of satellites. Fig.~\ref{fig.a3} shows the 
radial distribution of satellites for the \msg\ and \sfg\ samples obtained with the Millennium and M-XXL simulations. Here type 2 galaxies are shown
with pink shaded areas. Unlike the previous cases discussed before, the radial distribution of type 2 galaxies differs significantly between the
two simulation results. This is particularly important for the \msg\ sample, where the population of type 2 galaxies dominates over small distances. 
Interestingly, the \citet{guo11} model run over the Millennium simulation predicts a NFW-like profile for the distribution of satellites 
in the \msg\ sample, but is consistent with our finding for \sfg\ galaxies. 
The discrepancy in the radial distributions for the \msg\ sample arises in scales well below $1 \mpc$. Hence, this difference
is likely to affect the clustering in a small range of scales.

Furthermore, throughout this paper we tested different descriptions of intra-halo velocities by retaining everything else about the galaxy population. In particular, 
the radial distribution of galaxies in the galaxy samples was preserved among different galaxy catalogues with different intra-halo velocities. Therefore, our
conclusions about the impact of different intra-halo velocity descriptions are robust to resolution effects on the radial distribution of type 2 galaxies.

\bibliographystyle{mnras}
\bibliography{ref}

\bsp	
\label{lastpage}

\end{document}